\documentclass[a4paper,11pt]{article}
\usepackage{a4wide,amsmath,amssymb,bbm}
\usepackage[english]{babel}

\makeatletter

\@addtoreset{equation}{section} \makeatother

\begin{document}

\hfill{Imperial-TP-LW-2014-01}

\vspace{40pt}

\begin{center}
{\LARGE{\bf Superisometries and integrability of superstrings}}

\vspace{60pt}

Linus Wulff

\vspace{15pt}

{\it\small Blackett Laboratory, Imperial College, London SW7 2AZ, U.K.}\\

\vspace{120pt}

{\bf Abstract}

\end{center}
\noindent
For type II supergravity backgrounds with superisometries the corresponding transformations and conserved currents for the superstring are constructed up to fourth order in $\Theta$. It is then shown how, for certain backgrounds related to near horizon geometries of intersecting branes, the components of the superisometry current can be used to construct a Lax connection demonstrating the classical integrability of the string in these backgrounds. This includes examples of $AdS_2$ and $AdS_3$ backgrounds with a $D(2,1;\alpha)$ isometry group which have not previously been studied from an integrability point of view. The construction of the Lax connection is carried out up to second order in $\Theta$.

\pagebreak 
\tableofcontents
\setcounter{page}{1}

%%%%%%%%%%%%%%%%%%%%%%%%%%%%%%%%%%%%%%%%%%%%%%%%%%%%%%%%%%%%%%%%%%%%%%%%

\section{Introduction}
A major breakthrough in the study of the AdS/CFT-correspondence \cite{Maldacena:1997re} was the discovery of integrability, first on the field theory side \cite{Minahan:2002ve} and then on the string side \cite{Bena:2003wd}. Since then integrability techniques have been used to learn a lot about string theory in $AdS_5\times S^5$ and $\mathcal N=4$ super Yang-Mills, with the ultimate hope of eventually solving both theories and thereby proving the AdS/CFT-correspondence in this case (see \cite{Beisert:2010jr} for a comprehensive review).

A very interesting problem is to try to find more examples of AdS-backgrounds where the string is integrable (and their dual CFTs). One approach is to find deformations of the AdS$_5$/CFT$_4$-correspondence which preserve the integrability. Another approach, which we will take here, is to try to find other simple AdS-backgrounds, for example by considering the near-horizon geometries of intersecting branes, and demonstrate the (classical) integrability of the string by constructing a flat Lax connection. There are two known examples which have been studied a lot. The first one is string theory on $AdS_4\times\mathbbm{CP}^3$ with RR-flux \cite{Arutyunov:2008if,Stefanski:2008ik,Gomis:2008jt,Grassi:2009yj} which is dual to ABJM-theory \cite{Aharony:2008ug}. The second is string theory on $AdS_3\times S^3\times S^3\times S^1$ with RR and NSNS-flux \cite{Babichenko:2009dk,Cagnazzo:2012se} for which the dual 2d-CFT is not well understood. In both cases there is a way to fix the kappa symmetry of the string so that the string action reduces to a supercoset sigma model as in $AdS_5\times S^5$. The classical integrability is then easily demonstrated by the standard construction of a Lax connection \cite{Bena:2003wd}. The situation is not completely satisfactory however, since the kappa symmetry fixing employed is not consistent for all string configurations. An important example is the GKP-string rotating in $AdS_3$ whose low-energy limit has in both cases been shown to be integrable by other methods \cite{Bykov:2010tv,Basso:2012bw,Sundin:2013uca}\footnote{For $AdS_3\times S^3\times S^3\times S^1$ only the case with pure RR-flux was considered.}. We do not expect the integrability to depend on the particular string configuration considered so one expects the full string action, before fixing any kappa symmetry, to be integrable. This was indeed shown to be the case for strings in $AdS_4\times\mathbbm{CP}^3$ in \cite{Sorokin:2010wn} and for strings in $AdS_3\times S^3\times S^3\times S^1$ with pure RR-flux in \cite{Sundin:2012gc}. The construction of the Lax connection was only carried out to quadratic order in the fermions $\Theta$ due to the complexity of the calculations at higher orders\footnote{A reduced $AdS_4$-model was however shown to be integrable to all orders.}. The construction used components of the superisometry Noether current as building blocks for the Lax connection. The same construction has also been applied to strings in $AdS_2\times S^2\times T^6$ with RR-flux and the integrability has again been demonstrated to quadratic order in $\Theta$ \cite{Sorokin:2011rr,Cagnazzo:2011at}. Since this background preserves only eight supersymmetries the corresponding supercoset model is only a (classically consistent) truncation of the full string action and can not be directly used to argue the integrability. However, because the truncation to the supercoset does not commute with kappa symmetry, we expect/hope that kappa symmetry should be powerful enough to guarantee the integrability of the full string action. In this paper we will see that this expectation is indeed borne out in several examples.

Intersecting brane constructions give rise, by taking the near horizon limit (and dimensionally reducing if they are in eleven dimensions), to many simple examples of AdS-backgrounds for which one could hope that the string would be integrable. By simple we mean backgrounds with constant fluxes and dilaton. Several $AdS\times S\times S\times T$ backgrounds arising from intersecting branes were constructed in \cite{Boonstra:1998yu} using the intersection rules of \cite{Papadopoulos:1996uq,Tseytlin:1996bh,Gauntlett:1996pb,Tseytlin:1996hi}. They are listed in table \ref{table:1}.\footnote{We list only the type IIA solutions but all backgrounds except (A) can be trivially T-dualized along a toroidal direction to type IIB.} Backgrounds (C)--(G) appear to not have been studied in the literature before from an integrability point of view.\footnote{A special case of (G), $AdS_2\times S^2\times T^6$ with RR-flux, was studied in \cite{Sorokin:2011rr,Cagnazzo:2011at} as already remarked. The (non-critical) supercoset corresponding to $AdS_2\times S^2\times S^2$ (and a different realization of $AdS_2\times S^3$ to the one that appears here) occurs in the list of semi-symmetric supercosets with vanishing one-loop beta function given in \cite{Zarembo:2010sg} (the corresponding finite gap equations were discussed in \cite{Zarembo:2010yz}).
}

\begin{table}[ht]
\begin{tabular}{c|cccc}
& Space & Superisometry group & \#SUSYs & Parameters\\[10pt]
\hline & & &\\
A & $AdS_4\times CP^3$ & $OSp(6|4)$ & 24 & --\\[10pt]
B & $AdS_3\times S^3\times S^3\times S^1$ & $D(2,1;\alpha)^2\times U(1)$ & 16 & $0\leq\alpha\,,q\leq1$\\[10pt]
C & $AdS_3\times S^2\times S^3\times T^2$ & $D(2,1;\alpha)\times SL(2,\mathbbm{R})\times SU(2)\times U(1)^2$ & 8 & $0\leq\alpha\leq\infty$\\[10pt]
D & $AdS_3\times S^2\times S^2\times T^3$ & $D(2,1;\alpha)\times SL(2,\mathbbm{R})\times U(1)^3$ & 8 & $0\leq\alpha\leq1$\\[10pt]
E & $AdS_2\times S^3\times S^3\times T^2$ & $D(2,1;\alpha)\times SU(2)^2\times U(1)^2$ & 8 & $0\leq\alpha\leq1$\\[10pt]
F & $AdS_2\times S^2\times S^3\times T^3$ & $D(2,1;\alpha)\times SU(2)\times U(1)^3$ & 8 & $0\leq\alpha\leq\infty$\\[10pt]
G & $AdS_2\times S^2\times S^2\times T^4$ & $D(2,1;\alpha)\times U(1)^4$ & 8 & $0\leq\alpha\,,q\leq1$\\[10pt]
\end{tabular}
\caption{Integrable Type IIA string backgrounds arising from intersecting branes.}
\label{table:1}
\end{table}

Our aim in this paper is to show that the string is in fact (classically) integrable in all the backgrounds listed in table \ref{table:1} (before fixing kappa symmetry).\footnote{Since the bosonic background in all cases is a symmetric space the bosonic string should be integrable even with the NSNS flux. This is (probably) not enough for the full superstring to be integrable however, see the discussion in section \ref{sec:summary}.} This is done up to quadratic order in $\Theta$ by explicitly constructing the Lax connection using components of the superisometry current along the lines of \cite{Sorokin:2010wn,Sundin:2012gc,Sorokin:2011rr,Cagnazzo:2011at}. The construction is complicated somewhat by the fact that all backgrounds except (A) generically involve both RR and NSNS flux whereas the backgrounds studied previously, using this construction, only involved RR-flux. In table \ref{table:1} we have also listed the corresponding superisometry group, which in all cases except (A) involves the exceptional supergroup $D(2,1;\alpha)$ with $\alpha$ a real parameter (note that in examples (C)--(F) only the left $SL(2,\mathbbm{R})_L$ ($SU(2)_L$) of $AdS_3$ ($S^3$) sits inside $D(2,1;\alpha)$). This free parameter turns out to control the relative curvature radii of two factors in the geometry. Two examples are
\begin{eqnarray}
\text{(C):}&& R_{AdS_3}=\frac{2}{\sqrt{1+\alpha}}\,,\qquad R_{S^2}=1\,,\qquad R_{S^3}=\frac{2}{\sqrt\alpha}\nonumber\\
\text{(D):}&& R_{AdS_3}=2\,,\qquad R_{S^2_1}=\frac{1}{\sqrt\alpha}\,,\qquad R_{S^2_2}=\frac{1}{\sqrt{1-\alpha}}\,.
\label{eq:radius-rel}
\end{eqnarray}
In the limit $\alpha\rightarrow0$ (and sometimes $\alpha\rightarrow1$) an $S^2$ or $S^3$ decompactifies and one obtains a different geometry. For example taking $\alpha=0$ in (C) gives $AdS_3\times S^2\times T^5$. This and similar backgrounds are therefore special cases of the ones listed in table \ref{table:1}. Note also that taking $\alpha\rightarrow\infty$ in (C) gives a highly curved $AdS_3\times S^3$ part while the curvature of $S^2$ remains finite. For some purposes it may be more convenient to keep the $AdS$-radius fixed. Rescaling all radii by $\sqrt{1+\alpha}$ and introducing $\tilde\alpha=\frac{1}{1+\alpha}$ get instead
\begin{equation}
\text{(C): }\quad R_{AdS_3}=2\,,\qquad R_{S^2}=\frac{1}{\sqrt{\tilde\alpha}}\,,\qquad R_{S^3}=\frac{2}{\sqrt{1-\tilde\alpha}}\,,\nonumber\\
\end{equation}
and similarly for (F). Here $0<\tilde\alpha\leq1$ since for $\tilde\alpha=0$ the fluxes of the supergravity solution diverge. The superisometry algebra $D(2,1;\tilde\alpha)$ then takes exactly the same form as in (B), (D), (E) and (G). This explains what we mean by $D(2,1;\alpha)$ in (C) and (F).

Backgrounds (B) and (G) actually have one more free parameter that we call $0\leq q\leq1$. It controls the amount of NSNS flux with $q=0$ corresponding to pure RR flux (in (B) $q=1$ corresponds pure NSNS flux). This additional free parameter can be understood most easily in the dual type IIB picture where it arises from the freedom to perform an $SL(2,\mathbbm R)$ S-duality.

The outline of the paper is as follows. In section \ref{sec:superisometries} we give a general discussion of the form of the superisometry transformations in a type II supergravity background. We describe how the transformations may be determined order by order in $\Theta$ and go on to determine them up to order $\Theta^4$ using the results of \cite{Wulff:2013kga}. We also give the corresponding Noether currents for the superstring to the same order. The discussion in section \ref{sec:superisometries} is completely general but the only result needed for the integrability discussion in the rest of the paper is the form of the superisometry Noether current up to order $\Theta^2$. In the second part of the paper we focus on very special backgrounds, namely symmetric spaces with constant fluxes and constant dilaton. We postulate a general form for the superisometry algebra in section \ref{sec:algebra} and describe some additional conditions needed on the fluxes for integrability in the case of both RR and NSNS flux. We then go on to construct the Lax connection up to quadratic order in $\Theta$. Section \ref{sec:backgrounds} gives the details of the backgrounds listed in table \ref{table:1} and we show that they fulfill all the conditions needed for the construction of the Lax connection in the previous section. We end the paper with some conclusions.

\section{Strings in backgrounds with superisometries}\label{sec:superisometries}
For backgrounds with some isometries and which preserve some amount of supersymmetry the string action will of course be invariant under the corresponding transformations. To find the explicit form of the transformations of the coordinates $(x,\Theta)$ of superspace which leave the action invariant one needs to construct the Killing vector and Killing spinor superfields. We will now describe the general procedure for doing this. A similar discussion for maximally supersymmetric backgrounds can be found in \cite{Claus:1998yw}. 

\subsection{Isometries in superspace}
The infinitesimal transformation of the supercoordinates\footnote{The transformation can be expanded as $K^M=\epsilon^{\mathcal M}K_{\mathcal M}{}^M(x,\Theta)$, where the index $\mathcal M$ runs over the generators of the isometry group and $\epsilon^{\mathcal M}$ are constant infinitesimal parameters.}
\begin{equation}
\delta z^M=K^M(x,\Theta)\,,\qquad z^M=(x^m,\Theta^\mu)\,,
\label{eq:superisometry-tranf}
\end{equation}
where $K^M=(K^m\,,K^\mu)$ is some superfield, is a superisometry if the supervielbeins transform only by an induced Lorentz transformation, i.e.
\begin{equation}
\delta E^a=l^a{}_bE^b\,,\qquad \delta E^\alpha=\frac{1}{4}(\Gamma^{ab}E)^\alpha\,l_{ab}\,,
\label{eq:Eisometry}
\end{equation}
for some anti-symmetric matrix $l_{ab}$ which may depend on $x$ and $\Theta$. The fields of the supergravity background of course also have to respect the isometry, i.e. the dilaton, NSNS three-form and the RR field strengths must satisfy
\begin{equation}
\mathcal L_{\delta z}\phi=0\,,\qquad\mathcal L_{\delta z}H=0\,,\qquad\mathcal L_{\delta z}F^{(n)}=0\,.
\end{equation}
Plugging in $\delta z^M=K^M$ into (\ref{eq:Eisometry}) leads to equations for $K^M$. From the first equation we get
\begin{eqnarray}
l^a{}_bE^b
%=dK^a+E^C K^B (T_{BC}{}^a-2\Omega_{[BC]}{}^a)
=
dK^a+K^b\Omega_b{}^a
+E^C K^B (T_{BC}{}^a-\Omega_{BC}{}^a)
\,,\qquad K^a=K^ME_M{}^a\,,
\end{eqnarray}
where $K^a$ is the Killing vector superfield. Using the superspace torsion constraints of \cite{Wulff:2013kga} this becomes
\begin{equation}
\nabla_{(a}K_{b)}=0\qquad\mbox{and}\qquad\nabla_\alpha K^a-i(\Gamma^a\Xi)_\alpha=0\,,\qquad \Xi^\alpha=K^ME_M{}^\alpha\,,\label{eq:dalphaKa}
\end{equation}
where $\Xi$ is the Killing spinor superfield. We also get
\begin{equation}
l_{ab}=-\nabla_aK_b+K^C\Omega_{Cab}\,.
\end{equation}
Using $\delta z^M=K^M$ in the second equation of (\ref{eq:Eisometry}) gives
\begin{equation}
\frac{1}{4}(\Gamma^{ab}E)^\alpha\,l_{ab}
%=d\Xi^\alpha+E^C K^B (T_{BC}{}^\alpha-2\Omega_{[BC]}{}^\alpha)
=
d\Xi^\alpha+\Xi^\beta\Omega_\beta{}^\alpha
+E^CK^B(T_{BC}{}^\alpha-\Omega_{BC}{}^\alpha)\,,
\end{equation}
which, using the superspace torsion constraints of \cite{Wulff:2013kga}, gives the superfield Killing spinor equation
\begin{eqnarray}
\nabla_a\Xi^\alpha+\frac18\big([H_{abc}\Gamma^{bc}\Gamma_{11}+S\Gamma_a]\Xi\big)^\alpha-\psi_{ab}{}^\alpha\,K^b=0
\label{eq:Killingspinor}
\end{eqnarray}
and an equation for the spinor derivative of the Killing spinor superfield which determines the higher components in the $\Theta$-expansion
\begin{eqnarray}
&&\nabla_\beta\Xi^\alpha
-\frac18\big([H_{abc}\Gamma^{bc}\Gamma_{11}+S\Gamma_a]\big)^\alpha{}_\beta\,K^a
+\frac14(\Gamma^{ab})^\alpha{}_\beta\,\nabla_aK_b
-\frac12\chi_\beta\,\Xi^\alpha
+\frac12\delta^\alpha_\beta\,\Xi\chi
\label{eq:dalphaXi}
\\
&&{}
+\frac12(\Gamma_{11}\chi)_\beta\,(\Gamma_{11}\Xi)^\alpha
-\frac12(\Gamma_{11})^\alpha{}_\beta\,\Xi\Gamma_{11}\chi
-\frac12(\Gamma^a\Xi)_\beta\,(\Gamma_a\chi)^\alpha
+\frac12(\Gamma^a\Gamma_{11}\Xi)_\beta\,(\Gamma_a\Gamma_{11}\chi)^\alpha
=0\,.
\nonumber
\end{eqnarray}
Here $\psi_{ab}$ is the gravitino field strength superfield, $\chi$ is the dilatino superfield and $S$ is a superfield constructed from the RR field strengths contracted with gamma matrices \cite{Wulff:2013kga}. The bosonic part of $S$ is given, in the type IIA case, by
\begin{equation}
S=e^\phi\big(\frac12F_{ab}^{(2)}\Gamma^{ab}\Gamma_{11}+\frac{1}{4!}F_{abcd}^{(4)}\Gamma^{abcd}\big)\,.
\label{eq:S}
\end{equation}
The condition that the dilaton superfield respect the isometry gives the superspace dilatino equation
\begin{equation}
0=\mathcal L_{\delta z}\phi=K^M\partial_M\phi=\Xi\chi+K^a\nabla_a\phi\,.
\label{eq:superdilatino}
\end{equation}

To find the explicit form of the superisometry transformations one simply has to find the form of the superfields $K^a$ and $\Xi^\alpha$ and the form of the supervielbeins since
\begin{equation}
\delta z^ME_M{}^a=K^a\,,\qquad\delta z^ME_M{}^\alpha=\Xi^\alpha\,.
\end{equation}
The string action and the form of the supervielbeins for a general supergravity background is known up to order $\Theta^4$ \cite{Wulff:2013kga}. We will now determine $K^a$ and $\Xi^\alpha$ to this order as well (it is sufficient to know $\Xi$ to order $\Theta^3$).

\subsection{$\Theta$-expansion of the Killing vector and Killing spinor superfields}\label{sec:thetaexp}
The procedure is almost identical to the procedure for finding the supervielbeins order by order in $\Theta$ using the supergravity constraints and the bosonic geometry as input. One introduces a parameter $t$ and rescales $\Theta\rightarrow t\Theta$ in all superfields. Using the fact that\footnote{The Wess-Zumino gauge like condition $i_\Theta\Omega^{AB}=0$ is imposed on the spin connection while $i_\Theta E^\alpha=\Theta^\alpha$.} $\frac{d}{dt}=\Theta^\alpha\nabla_\alpha$ when acting on a superfield one uses the superspace constraints to write first order ordinary differential equations for the $t$-dependence, i.e. $\Theta$-dependence, of the relevant superfields. Using (\ref{eq:dalphaKa}) and  (\ref{eq:dalphaXi}) we find the equations\footnote{Throughout this section we will write the expressions relevant to type IIA supergravity, i.e. $\Theta$ is a 32-component Majorana spinor. However, with very minor changes they also hold for type IIB, see \cite{Wulff:2013kga}.
}
\begin{align}
\frac{d}{dt}K^a&=i\Theta\Gamma^a\Xi\,,\label{eq:ddtK}\allowdisplaybreaks\\
\frac{d}{dt}\Xi^\alpha&=
-\frac14(\Gamma^{ab}\Theta)^\alpha\,\nabla_aK_b
+\frac18\big([H_{abc}\Gamma^{bc}\Gamma_{11}+S\Gamma_a]\Theta\big)^\alpha\,K^a
-\frac12\Theta^\alpha\,\Xi\chi
+\frac12(\Gamma_{11}\Theta)^\alpha\,\Xi\Gamma_{11}\chi
\nonumber\\
&\qquad
+\frac12\Xi^\alpha\,\Theta\chi
-\frac12(\Gamma_{11}\Xi)^\alpha\Theta\Gamma_{11}\chi
+\frac12(\Gamma_a\chi)^\alpha\,\Theta\Gamma^a\Xi
-\frac12(\Gamma_a\Gamma_{11}\chi)^\alpha\,\Theta\Gamma^a\Gamma_{11}\Xi\,,
\label{eq:ddtXi}\allowdisplaybreaks\\
\frac{d}{dt}\nabla_aK_b&=\Theta^\alpha\nabla_\alpha\nabla_aK_b
=
-\Theta^\alpha T_{\alpha a}{}^C\nabla_CK_b
-\Theta^\alpha R_{\alpha abc}K^c
+i\Theta\Gamma_b\nabla_a\Xi
\nonumber\\
&=
\frac{i}{4}\Theta\Gamma_{[a}S\Gamma_{b]}\Xi
-\frac{i}{2}H_{abc}\,\Theta\Gamma^c\Gamma_{11}\Xi
-i\Theta\Gamma_{[a}\psi_{b]c}\,K^c
+\frac{i}{2}\Theta\Gamma_c\psi_{ab}\,K^c\,.
\label{eq:ddtnablaK}
\end{align}
These equations are in fact identical to the ones for the supervielbeins and spin connection written in \cite{Wulff:2013kga} with the replacements
\begin{equation}
E^a\rightarrow K^a\,,\qquad E^\alpha\rightarrow\Xi^\alpha\,,\qquad\Omega_{ab}\rightarrow\nabla_aK_b\,,\qquad d\Theta\rightarrow0\,.
\end{equation}
These equations can now be solved order by order in $\Theta$ subject to the boundary conditions that at $\Theta=0$ we should have
\begin{equation}
\Xi^{(0)}=\xi\,,\qquad K_a^{(0)}=k_a\,,\qquad (\nabla_aK_b)^{(0)}=\nabla_ak_b\,,
\end{equation}
where $k_a(x)$ is the Killing vector and $\xi(x)$ is the Killing spinor. Note that in the last equation the covariant derivative on the left-hand-side involves the full spin connection superfield while on the right-hand-side it involves only the spin connection of the bosonic background $\omega^{ab}(x)$.
The Killing spinor $\xi$ satisfies the Killing spinor equation, which can be obtained by setting $\Theta=0$ in the corresponding superfield equation (\ref{eq:Killingspinor}),
\begin{equation}
\mathcal D_a\xi=\nabla_a\xi+\frac18M_a\xi=0\qquad\text{where}\qquad M_a=H_{abc}\Gamma^{bc}\Gamma_{11}+S\Gamma_a\,.
\label{eq:Killing}
\end{equation}
The integrability condition for the Killing spinor equation is
\begin{equation}
U_{ab}\xi=0\qquad\text{where}\qquad U_{ab}=-\frac14R_{ab}{}^{cd}\Gamma_{cd}+\frac{1}{32}M_{[a}M_{b]}+\frac14\nabla_{[a}M_{b]}\,,
\end{equation}
which is also the condition for supersymmetry coming from the variation of the gravitino \cite{Wulff:2013kga}.

Evaluating (\ref{eq:ddtK}), (\ref{eq:ddtXi}) and (\ref{eq:ddtnablaK}) at $t=0$, using the fact that all fermionic fields except $\Xi$ vanish at lowest order in $\Theta$, we find at the linear order in $\Theta$
\begin{eqnarray}
&&K_a^{(1)}=i\Theta\Gamma_a\xi\,,\qquad(\nabla_aK_b)^{(1)}=\frac{i}{8}\Theta\Gamma_{[a}M_{b]}\xi-\frac{i}{8}\xi\Gamma_{[a}M_{b]}\Theta
%=\frac{i}{4}\Theta\Gamma_{[a}S\Gamma_{b]}\xi-\frac{i}{2}H_{abc}\,\Theta\Gamma^c\Gamma_{11}\xi
\,,
\nonumber\\
&&\Xi^{(1)}=\frac18(M^a\Theta)\,k_a-\frac14(\Gamma^{ab}\Theta)\,\nabla_ak_b\,.
\end{eqnarray}
Applying a derivative to (\ref{eq:superdilatino}) we also get the dilatino equation
\begin{equation}
0=\Xi T\Theta+i\nabla_a\phi\,\Theta\Gamma^a\xi=\Theta T\xi\qquad\Rightarrow\qquad T\xi=0\,,
\label{eq:dilatino}
\end{equation}
where
\begin{equation}
T=\frac{i}{2}\nabla_a\phi\Gamma^a+\frac{i}{24}H_{abc}\Gamma^{abc}\Gamma_{11}+\frac{i}{16}\Gamma^aS\Gamma_a\,.
\end{equation}

Applying another $t$-derivative to (\ref{eq:ddtK}), (\ref{eq:ddtXi}) and (\ref{eq:ddtnablaK}) and evaluating at $t=0$ using $E^{(1)}=\mathcal D\Theta=\nabla\Theta+\frac18e^aM_a\Theta$, $\psi_{ab}^{(1)}=U_{ab}\Theta$ and $\chi^{(1)}=T\Theta$ (see \cite{Wulff:2013kga}) we find at the second order in $\Theta$
\begin{align}
K_a^{(2)}&=\frac{i}{2}\Theta\Gamma_a\Xi^{(1)}=
\frac{i}{16}\Theta\Gamma_aM^b\Theta\,k_b
-\frac{i}{8}\Theta\Gamma_a{}^{bc}\Theta\,\nabla_bk_c\,,
\nonumber\\
(\nabla_aK_b)^{(2)}&=
\frac{i}{16}\Theta\Gamma_{[a}M_{b]}\Xi^{(1)}
-\frac{i}{16}\Xi^{(1)}\Gamma_{[a}M_{b]}\Theta
-\frac{i}{2}\Theta\Gamma_{[a}U_{b]c}\Theta\,k^c
+\frac{i}{4}\Theta\Gamma_cU_{ab}\Theta\,k^c\,,
\nonumber\\
\Xi^{(2)}&=\frac12\mathcal M\xi-\frac14(M+\tilde M)\xi\,,
\end{align}
where we have introduced the matrices
\begin{align}
\mathcal M^\alpha{}_\beta&=
M^\alpha{}_\beta
+\tilde M^\alpha{}_\beta
+\frac{i}{8}(M_a\Theta)^\alpha\,(\Theta\Gamma^a)_\beta
-\frac{i}{32}(\Gamma^{ab}\Theta)^\alpha\,(\Theta\Gamma_aM_b)_\beta
-\frac{i}{32}(\Gamma^{ab}\Theta)^\alpha\,(C\Gamma_aM_b\Theta)_\beta\,,
\nonumber\\
M^\alpha{}_\beta&=
\frac12\Theta T\Theta\,\delta^\alpha_\beta
-\frac12\Theta\Gamma_{11}T\Theta\,(\Gamma_{11})^\alpha{}_\beta
+\Theta^\alpha\,(CT\Theta)_\beta
+(\Gamma^aT\Theta)^\alpha\,(\Theta\Gamma_a)_\beta\,,\qquad \tilde M=\Gamma_{11}M\Gamma_{11}\,,\nonumber
\end{align}
which also appear, written in a slightly different way, in \cite{Wulff:2013kga}. In the above expressions $C$ is the charge-conjugation matrix and we follow the conventions of \cite{Wulff:2013kga}.

Continuing to the next order we find, using the lower order results and the expressions given in \cite{Wulff:2013kga} for $H^{(2)}_{abc}$ and $S^{(2)}$,
\begin{align}
K_a^{(3)}&=\frac{i}{3}\Theta\Gamma_a\Xi^{(2)}=\frac{i}{6}\Theta\Gamma_a\mathcal M\xi-\frac{i}{12}\Theta\Gamma_a(M+\tilde M)\xi\,,
\\
\Xi^{(3)}&=\frac16(\mathcal M\Xi^{(1)})^\alpha
+\frac{1}{96}\big([M+\tilde M]S\Gamma^a\Theta\big)^\alpha\,k_a
+\frac{1}{96}\big(\Theta\Gamma^a[M+\tilde M]SC\big)^\alpha\,k_a
\nonumber\\
&\qquad
-\frac{i}{24}(\Gamma^{ab}\Theta)^\alpha\,\Theta\Gamma^cU_{ab}\Theta\,k_c
+\frac{i}{24}(\Gamma^{ab}\Gamma_{11}\Theta)^\alpha\,\Theta\Gamma^c\Gamma_{11}U_{ab}\Theta\,k_c
+\frac{i}{24}(\Gamma^{ab}\Theta)^\alpha\,\Theta\Gamma_aU_{bc}\Theta\,k^c
\nonumber\\
&\qquad
+\frac{i}{24}(\Gamma^{ab}\Gamma_{11}\Theta)^\alpha\,\Theta\Gamma_a\Gamma_{11}U_{bc}\Theta\,k^c
+\frac{i}{48}\Theta\Gamma^{abc}\Gamma_{11}\Theta\,(\Gamma_{11}U_{ab}\Theta)^\alpha\,k_c
-\frac{i}{48}\Theta\Gamma^{abc}\Theta\,(U_{ab}\Theta)^\alpha\,k_c\,,
\nonumber
%\\
%(\nabla_aK_b)^{(3)}&=&
%\frac{i}{12}\Theta\Gamma_{[a}S\Gamma_{b]}\Xi^{(2)}
%+\frac{i}{12}\Theta\Gamma_{[a}S^{(2)}\Gamma_{b]}\xi
%-\frac{i}{6}H^{(2)}_{abc}\,\Theta\Gamma^c\Gamma_{11}\Xi
%-\frac{i}{6}H_{abc}\,\Theta\Gamma^c\Gamma_{11}\Xi^{(2)}
%-\frac{i}{3}\Theta\Gamma_{[a}U_{b]c}\Theta\,K^{(1)c}
%+\frac{i}{6}\Theta\Gamma_cU_{ab}\Theta\,K^{(1)c}
\end{align}
We have left out $(\nabla_aK_b)^{(3)}$ since it is only needed at the next order in $\Theta$. And finally we have
\begin{equation}
K_a^{(4)}=\frac{i}{4}\Theta\Gamma_a\Xi^{(3)}\,.
\end{equation}

Note that $K^{(2)}_a$, $(\nabla_aK_b)^{(2)}$ and $\Xi^{(3)}$ can also be obtained directly by making the replacements
\begin{equation}
\mathcal D\Theta\rightarrow\Xi^{(1)}\qquad\text{and}\qquad e^a\rightarrow k^a
\end{equation}
in the expressions given in \cite{Wulff:2013kga} for $E^{(2)\,a}$, $\Omega^{(2)\,ab}$ and $E^{(3)\,\alpha}$ respectively.

\subsection{Worldsheet superisometry Noether current}
The Green--Schwarz string action in a general supergravity background takes the form
\begin{equation}
S=-T\int_\Sigma\,\left(\frac12*E^aE^b\eta_{ab}-B\right)\,,
\end{equation}
where $B$ is (the worldsheet pullback of) the NSNS two-form potential, $H=dB$. A star denotes the worldsheet Hodge dual defined with the (auxiliary) worldsheet metric. Using the superspace constraints of \cite{Wulff:2013kga} the equations of motion read
\begin{equation}
\nabla*E^a-\frac{i}{2}E\Gamma^a\Gamma_{11}E+\frac12E^cE^bH^a{}_{bc}=0\,,\qquad*E^a\,(\Gamma_aE)^\alpha-E^a\,(\Gamma_a\Gamma_{11}E)^\alpha=0\,.
\label{eq:eom}
\end{equation}
The superisometry Noether current takes the form
\begin{equation}
J=E^ai_{\delta z}E_a-*i_{\delta z}B+*\Lambda=E^a\,K_a-*i_{\delta z}B+*\Lambda\,,
\label{eq:Jprime}
\end{equation}
where $\delta z$ is the superisometry transformation of the supercoordinates given in (\ref{eq:superisometry-tranf}) and $B$ transforms by a gauge-transformation, $\delta B=d\Lambda$, leaving $H=dB$ invariant. Indeed we find, using the from of $\nabla K_a$ given in (\ref{eq:dalphaKa}),
\begin{eqnarray}
d*J&=&\nabla*E^a\,K_a+*E^a\,\nabla K_a-di_{\delta z}B+d\Lambda
\\
&=&
(\nabla*E_a-\frac{i}{2}E\Gamma_a\Gamma_{11}E+\frac12E^cE^bH_{abc})K^a
-i(*E^a\,\Xi\Gamma_aE-E^a\,\Xi\Gamma_a\Gamma_{11}E)
-\mathcal L_{\delta z}B
+d\Lambda\,.\nonumber
\end{eqnarray}
The first and second term are proportional to the bosonic and fermionic equation of motion respectively and the last two terms cancel since the change of $B$ under an isometry transformation is $d\Lambda=\delta B=\mathcal L_{\delta z}B$. This proves that $J$ is conserved on--shell.

To have a completely explicit form of $J$ we need to determine the superfield one-form $\Lambda$. This is again easily done order by order in $\Theta$. Using the fact that
\begin{eqnarray}
d\Lambda=\delta B=\mathcal L_{\delta z}B=i_{\delta z}H+di_{\delta z}B\,,
\end{eqnarray}
we can write, using the superspace constraint on $H$,\footnote{This is the same trick that was used in \cite{Wulff:2013kga} to compute $B$ from $H$.}
\begin{eqnarray}
\Lambda&=&i_{\delta z}B+\lambda+\int_0^1dt\,i_\Theta i_{\delta z}H(x,t\Theta)
\nonumber\\
&=&i_{\delta z}B+\lambda-i\int_0^1dt\,\left(E^a(t\Theta)\,\Theta\Gamma_a\Gamma_{11}\Xi(t\Theta)-\Theta\Gamma^a\Gamma_{11}E(t\Theta)\,K_a(t\Theta)\right)\,,
\end{eqnarray}
where the lowest component of $\Lambda$, $\lambda(x)$, satisfies 
\begin{equation}
d\lambda=i_kH^{(0)}=\frac12e^be^aH^{(0)}_{abc}k^c\,.
\end{equation}
Using this expression for $\Lambda$ in (\ref{eq:Jprime}) the superisometry current takes the explicit form
\begin{eqnarray}
J=E^a\,K_a+*\lambda-i\int_0^1dt\,\left(*E^a\,\Theta\Gamma_a\Gamma_{11}\Xi-\Theta\Gamma^a\Gamma_{11}*E\,K_a\right)\,,
\label{eq:J}
\end{eqnarray}
which is straightforwardly evaluated using the $\Theta$-expansion of the supervielbeins and the Killing vector and Killing spinor superfields. Using the $\Theta$-expansions of $\Xi$ and $K_a$ derived in the previous section and the expansion of the supervielbeins derived in \cite{Wulff:2013kga} we find the following $\Theta$-expansion of the superisometry Noether current up to fourth order in $\Theta$
\begin{align}
J^{(0)}={}&e^a\,k_a+*\lambda\,,\qquad J^{(1)}=ie^a\,\Theta\Gamma_a\xi-i*e^a\,\Theta\Gamma_a\Gamma_{11}\xi\,,\label{eq:J01}
\\
\allowdisplaybreaks
&\nonumber\\
J^{(2)}={}&
\frac{i}{2}\Theta\Gamma^a\mathcal D\Theta\,k_a
+\frac{i}{2}\Theta\Gamma^a\Gamma_{11}*\mathcal D\Theta\,k_a
+\frac{i}{16}e^a\,\Theta\Gamma_aM^b\Theta\,k_b
\label{eq:J2}
\\
&-\frac{i}{16}*e^a\,\Theta\Gamma_a\Gamma_{11}M^b\Theta\,k_b
-\frac{i}{8}e^a\,\Theta\Gamma_a{}^{bc}\Theta\,\nabla_bk_c
+\frac{i}{8}*e^a\,\Theta\Gamma_a{}^{bc}\Gamma_{11}\Theta\,\nabla_bk_c\,,
\allowdisplaybreaks
\nonumber\\
&\nonumber\\
J^{(3)}={}&
\frac{i}{6}e^a\,\Theta\Gamma_a\mathcal M\xi
-\frac{i}{12}e^a\,\Theta\Gamma_a(M+\tilde M)\xi
-\frac{i}{6}*e^a\,\Theta\Gamma_a\Gamma_{11}\mathcal M\xi
\\
&+\frac{i}{12}*e^a\,\Theta\Gamma_a\Gamma_{11}(M+\tilde M)\xi
-\frac12\Theta\Gamma^a\mathcal D\Theta\,\Theta\Gamma_a\xi
+\frac16\Theta\Gamma^a*\mathcal D\Theta\,\Theta\Gamma_a\Gamma_{11}\xi
\nonumber\\
&-\frac13\Theta\Gamma^a\Gamma_{11}*\mathcal D\Theta\,\Theta\Gamma_a\xi\,,
\allowdisplaybreaks
\nonumber\\
&\nonumber\\
J^{(4)}={}&
\frac{i}{4}\Theta\Gamma^aE^{(3)}\,k_a
+\frac{i}{4}\Theta\Gamma^a\Gamma_{11}*E^{(3)}\,k_a
+\frac{i}{4}e^a\,\Theta\Gamma_a\Xi^{(3)}
-\frac{i}{4}*e^a\,\Theta\Gamma_a\Gamma_{11}\Xi^{(3)}
\\
&-\frac14\Theta\Gamma^a\mathcal D\Theta\,\Theta\Gamma_a\Xi^{(1)}
+\frac18\Theta\Gamma^a*\mathcal D\Theta\,\Theta\Gamma_a\Gamma_{11}\Xi^{(1)}
-\frac18\Theta\Gamma^a\Gamma_{11}*\mathcal D\Theta\,\Theta\Gamma_a\Xi^{(1)}\,.
\nonumber
\end{align}
We have chosen not to expand $J^{(4)}$ all the way due to the length of the resulting expression. In the following we will use (pieces of) $J^{(0)}$, $J^{(1)}$ and $J^{(2)}$ to build a Lax connection for the string in certain symmetric space backgrounds. To extend this calculation beyond the quadratic order in $\Theta$ one would need also the higher components of $J$.

\section{Integrability of the string in certain backgrounds}\label{sec:integrability}
So far the discussion has been valid for a general supergravity background but from now on we will specify to very special backgrounds for which we can demonstrate integrability. In particular we will require the bosonic background to be a symmetric space. We will also require the fluxes and dilaton to be (covariantly) constant (as we will see below we will also need some additional conditions on the form of the fluxes). Note that the backgrounds listed in table \ref{table:1} satisfy these conditions.

Since the bosonic backgrounds are symmetric spaces we expect the bosonic part of the string to be integrable. The first step is to write down the bosonic Lax connection $L^{(0)}$. Our problem is then to try to extend this Lax connection by fermionic terms up to order $\Theta^2$. In principle one could try to go to higher order in $\Theta$ but we do not attempt this here for two reasons: (i) the complexity of the expressions quickly become rather daunting, and (ii) we don't expect any obstructions to occur beyond the quadratic order. One may ask whether any obstructions can occur at all or whether it is always possible to extend a bosonic Lax connection to all orders in fermions. An argument for this would be that kappa symmetry relates the bosons and fermions of the string to each other. However, one can argue that this is unlikely to be true. For example, if we take one of the solutions which we have found to be integrable and flip the signs of some of the fluxes we still have a supergravity solution, but now it will generically break all the supersymmetries. The bosonic Lax connection will be the same but when we try to extend it to quadratic order we will see that we rely in a crucial way on having Killing spinors, see section \ref{sec:summary} (this argument was first given in \cite{Sorokin:2010wn} for the case of $AdS_4\times\mathbbm{CP}^3$). It therefore appears that it will generically not be possible to extend a bosonic Lax connection to quadratic order in fermions.\footnote{Note that there are certainly examples of backgrounds without supersymmetry in which the string should be integrable. A simple example is type IIA on $AdS_5\times\mathbbm{CP}^2\times S^1$ \cite{Duff:1998us} which is T-dual to type IIB on $AdS_5\times S^5$. Another example is the $\gamma$-deformation of $AdS_5\times S^5$ \cite{Frolov:2005dj}.} An interesting question is whether it is always possible given some minimal amount of supersymmetry.

We will start by describing our approach to constructing the Lax connection in a formal way as a deformation problem. Then we will postulate a form of the superisometry algebra and discuss the extra conditions on the fluxes that we will need before proceeding to the actual construction of the Lax connection. We will end the section with a brief summary and discussion.

\subsection{Deformation problem for the Lax connection}\label{sec:deformation}
The question of whether the classical integrability of the bosonic string in a certain background extends to the full superstring amounts to finding a deformation of the bosonic Lax connection by terms involving the fermions while preserving its flatness. The integrability of the bosonic sector implies that there exists a Lax connection $L^{(0)}$, independent of $\Theta$, such that it is flat modulo terms involving $\Theta$
\begin{equation}
dL^{(0)}-L^{(0)}L^{(0)}=\mathcal O(\Theta^2)\,.
\end{equation}
Note that in the above equation the full superstring equations of motion are used. Next we want to find a Lax connection linear in $\Theta$, $L^{(1)}$ (since the Lax connection is a bosonic object this will have to involve the Killing spinors). It must satisfy
\begin{equation}
DL^{(1)}=dL^{(1)}-L^{(0)}L^{(1)}-L^{(1)}L^{(0)}=\mathcal O(\Theta^3)\,.
\end{equation}
We then have a Lax connection up to linear order in $\Theta$ since
\begin{equation}
d(L^{(0)}+L^{(1)})-(L^{(0)}+L^{(1)})(L^{(0)}+L^{(1)})=\mathcal O(\Theta^2)\,.
\end{equation}
Let us call the order $\Theta^2$ terms on the right-hand-side $F^{(2)}$, i.e. the right-hand-side is $F^{(2)}+\mathcal O(\Theta^3)$. The Lax connection can be extended to quadratic order in $\Theta$ if and only if it is possible to write 
\begin{equation}
F^{(2)}=DL^{(2)}+\mathcal O(\Theta^4)\,.
\end{equation}
If so it is easy to see that $L=L^{(0)}+L^{(1)}+L^{(2)}$ is flat modulo terms cubic in $\Theta$. One can then go to the next order and so on. Note that it is very important for this process to work that one writes the most general possible form of $L^{(0)}$ and $L^{(1)}$ since otherwise one might mistakenly conclude that $L^{(2)}$ does not exist. We will see that this will be important in our case.

\subsection{Superisometry algebra}\label{sec:algebra}
Since the Lax connection is valued in the superisometry algebra we need to know the form of the latter before discussing the possible integrability. Since we are assuming that the bosonic geometry is a symmetric space the killing vectors should satisfy the standard algebra appropriate to a symmetric space, i.e.
\begin{equation}
[k_a,k_b]=\nabla_ak_b\,,\qquad[k_c,\nabla_ak_b]=R_{abc}{}^dk_d\,,
\label{eq:kkcom}
\end{equation}
where $R_{ab}{}^{cd}$ is the Riemann curvature. Note that this means that the Killing vector acts like the covariant derivative. For the commutator of the Killing vector with the Killing spinor it is natural to ask that the Killing vector should again act as a derivation and using the Killing spinor equation (\ref{eq:Killing}) we get
\begin{eqnarray}
[k_a,\xi]=\nabla_a\xi=-\frac18M_a\xi\,.
\end{eqnarray}
We will find however that we need the existence of a matrix $\hat S$ such that\footnote{Note that the fluxes in $M_a$ are assumed to be constant and therefore $\hat S$ is constant as well.}
\begin{equation}
M_a\xi=c_{[a]}\hat S\Gamma_a\xi\,.
\end{equation}
Note that when there is no NSNS flux this is trivial since in that case $M_a=S\Gamma_a$ so that $\hat S=S$. Here $c_{[a]}$ is either $1$ or $2$ depending on the index $a$. Writing $a=(\hat a,\,\tilde a,\,a')$, where $\hat a$ runs over any $AdS_2$ and $S^2$ directions $a'$ runs over the flat (i.e. toroidal) directions and $\tilde a$ runs over the rest, we have
\begin{equation}
c_{[\hat a]}=2\,,\qquad c_{[\tilde a]}=c_{[a']}=1\,.
\label{eq:ca}
\end{equation}
This factor comes from the difference in a factor of two between the $AdS_3$ ($S^3$) and $AdS_2$ ($S^2$) curvature radius, see for example (\ref{eq:radius-rel}). We therefore require the Killing vector -- Killing spinor commutator to take the form
\begin{equation}
[k_a,\xi]=\nabla_a\xi=-\frac{c_{[a]}}{8}\hat S\Gamma_a\xi\,.
\label{eq:kxi-comm}
\end{equation}
Note that for consistency $\hat S$ must satisfy the projection equation $\mathcal P\hat S=\hat S$ where $\mathcal P$ is the Killing spinor projector, i.e. $\mathcal P\xi=\xi$ (see section \ref{sec:backgrounds}). From these equations it also follows that
\begin{eqnarray}
[\nabla_ak_b,\xi]
=
\frac{c_{[a]}c_{[b]}}{32}\hat S\Gamma_{[b}\hat S\Gamma_{a]}\xi
=
-\frac14R_{ab}{}^{cd}\,\Gamma_{cd}\xi\,.
\label{eq:nablakxi-comm}
\end{eqnarray}

Finally we come to the commutator of two Killing spinors. We will simply postulate that it take the following form
\begin{equation}
\{\xi^\alpha,\xi^\beta\}=
-\frac{i}{8}(\hat S\Gamma^a\hat SC)^{\alpha\beta}\,k_a
+\frac{i}{4c_{[a]}}(\Gamma^{ab}\hat SC)^{\alpha\beta}\,\nabla_ak_b\,,
\label{eq:xixi-comm}
\end{equation}
where $C$ is the charge-conjugation matrix. For consistency we require also that $\hat S$ should satisfy the following conditions
\begin{align*}
	\text{(i)}\quad&(\hat SC)^{(\alpha\beta)}=0\\
	\text{(ii)}\quad&[\hat S,\Gamma^{ab}]\nabla_ak_b=0\\
	\text{(iii)}\quad&M_{a'}\xi=\hat S\Gamma_{a'}\xi=0\,.
\end{align*}
The first two conditions ensure that the right-hand-side of (\ref{eq:xixi-comm}) is indeed symmetric in the spinor indices. The last condition means that the Killing vectors of the toroidal directions $k_{a'}$ decouple from the algebra, see (\ref{eq:kxi-comm}). Note that these are constant, i.e. $\nabla k_{a'}=0$.

Since we have simply postulated the form of the commutators involving the killing spinors we have to ensure that the Jacobi identities are satisfied. Applying a covariant derivative $\nabla_c$ to (\ref{eq:xixi-comm}) and using (\ref{eq:kxi-comm}) and (\ref{eq:kkcom}) we find that
\begin{equation}
\frac{ic_{[c]}}{32}(\hat S\Gamma^c\hat S\Gamma^d\hat SC)^{(\alpha\beta)}\,k_d
-\frac{i}{4c_{[a]}}(\hat S\Gamma^{ab}C)^{\alpha\beta}\,R_{ab}{}^{cd}k_d
\end{equation}
should vanish. This is indeed true as can be seen by using (\ref{eq:nablakxi-comm}) and the symmetry properties of $\hat S$ and the gamma-matrices. This calculation is equivalent to checking the Jacobi identity involving one $k_a$ and two $\xi$. The last step that remains is to check the Jacobi identity involving three $\xi$. We find
\begin{equation}
0=[\{\xi^{(\alpha},\xi^\beta\},\xi^{\gamma)}]=
\frac{i}{128}[2c_{[a]}(\hat S\Gamma^a\hat SC)^{(\alpha\beta}(\hat S\Gamma_a\xi)^{\gamma)}-c_{[a]}(\hat S\Gamma^{ab}C)^{(\alpha\beta}(\hat S\Gamma_a\hat S\Gamma_b\xi)^{\gamma)}]\,.
\end{equation}
This Jacobi identity can be verified on a case by case basis for the backgrounds in table \ref{table:1} by making use of the properties of $\hat S$ and Fierz identities. We have not found a simple way to show that it holds in general. When there is no NSNS flux, so that $\hat S=S$, it is however easy to check (note that $S$ is anti-symmetric and commutes with $\Gamma_{11}$ by (\ref{eq:S})).

In addition to the constraints on $\hat S$ and the form of the superisometry algebra mentioned so far we find that we need some further constraints on the fluxes to be satisfied for our Lax connection construction to go through. These extra conditions are related to the fact that we have both RR and NSNS flux. When there is no NSNS flux they become trivial. We divide the possible backgrounds into to groups: (I) The ones without $AdS_2$ and $S^2$ factors ((A) and (B) in table \ref{table:1}) and (II) the ones with at least one $AdS_2$ or $S^2$ factor ((C)--(G) in table \ref{table:1}). The conditions we impose are listed in table \ref{table:2} and \ref{table:3} respectively. The first condition says that for group I the NSNS flux is only allowed to be on $AdS_3$ or $S^3$ while for group II it is only allowed to be on $AdS_2\times\mathbbm R$ or $S^2\times\mathbbm R$ where $\mathbbm R$ denotes one of the toroidal (flat) directions. In particular the two types of NSNS flux cannot coexist. The second condition states that the NSNS flux is distributed evenly between the $AdS_{2,3}$ and $S^{2,3}$ factors in the geometry with the parameter $q$ measuring the amount of flux. Note that the Riemann curvature of $AdS$ ($S$) is simply $R_{ab}{}^{cd}=+(-)\frac{2}{R^2}\delta_{[a}^c\delta_{b]}^d$ so that $h_{abc}$ essentially coincides with $H_{abc}$ except for signs and factors of the curvature radius $R$. For the remaining conditions we need to assume that, if there is non-zero NSNS flux, there is at least one toroidal (flat) direction that we label by $9$. We then define the ''rotated'' gamma matrices
\begin{equation}
\Gamma_{9'}=\hat q\Gamma^9-q\Gamma_{11}\,,\qquad \Gamma_{11'}=q\Gamma^9+\hat q\Gamma_{11}\qquad\text{where}\qquad q^2+\hat q^2=1\,.
\label{eq:rotatedgamma}
\end{equation}
Note that this ''rotation'' preserves the Clifford algebra of the gamma matrices. Conditions (vi), (vii) and (viii) and conditions (vi$'$), (vii$'$) and (ix$'$) give certain conditions on $\hat S$ and $M_a$, appearing on the right-hand-side of the Killing spinor equation (\ref{eq:Killing}), involving these gamma matrices. Condition (ix) and (x$'$) expresses $S$ in terms if $\hat S$. Conditions (x), (xi) and (xi$'$) relates certain products of $S$, $\hat S$ and gamma matrices to the NSNS flux. Finally condition (viii$'$) says that for group II $q\neq1$ only if there are only $AdS_2$ and $S^2$ factors in the geometry (plus flat directions of course). This corresponds to (G) in our list in table \ref{table:1}. Note that for condition (xii$'$) to make sense we have to assume that $AdS_2$ or $S^2$ factors can only occur in the geometry together with $AdS_3$ or $S^3$ factors (or with other $AdS_2$ or $S^2$ factors) so that the $\varepsilon$-symbol is defined ($R_{\tilde a}$ denotes the corresponding curvature radius). Again this is clearly true for the backgrounds listed in table \ref{table:1}.

\begin{table}[ht]
\begin{align*}
	\text{(iv)}\quad&H_{abc}=H_{\tilde a\tilde b\tilde c}\quad\text{with}\quad\tilde a\,,\tilde b\,,\tilde c\in AdS_3(S^3)\\
  \text{(v)}\quad&h_{\tilde a\tilde b\tilde c}H^{\tilde c\tilde d\tilde e}=-4q^2\delta_{[\tilde a}^{\tilde c}\delta_{\tilde b]}^{\tilde d}\quad\text{where}\quad R_{ab}{}^{de}h_{dec}=H_{abc}\\
	\text{(vi)}\quad&[M_a,q\Gamma_{9'}]\xi=0\\
	\text{(vii)}\quad&\{\hat S,q\Gamma_{9'}\}=0\\
	\text{(viii)}\quad&q\hat q\hat S\Gamma_{9'}=qS\Gamma^9\\
	\text{(ix)}\quad&2S=\hat S+\Gamma_{11}\hat S\Gamma_{11}\\
	\text{(x)}\quad&S\Gamma^a(\hat S-\Gamma_{11}\hat S\Gamma_{11})=-2H^{abc}S\Gamma_{bc}\Gamma_{11}\\
  \text{(xi)}\quad&qS\Gamma_{\tilde a9}S=-\hat qH_{\tilde a\tilde b\tilde c}\Gamma^{\tilde b\tilde c}S
\end{align*}
\caption{Group I: Conditions for backgrounds without $AdS_2$ or $S^2$ factors.}
\label{table:2}
\end{table}

\begin{table}[ht]
\begin{align*}
	\text{(iv$'$)}\quad&H_{abc}=H_{\hat a\hat bc'}\qquad\hat a\,,\hat b\in AdS_2(S^2)\\
	\text{(v$'$)}\quad&h_{\hat a\hat be'}H^{e'\hat c\hat d}=-q^2\delta_{[\hat a}^{\hat c}\delta_{\hat b]}^{\hat d}\,,\quad h_{a'\hat b\hat e}H^{\hat e\hat cd'}=\frac12q^2\delta_{a'}^{d'}\delta_{\hat b}^{\hat c}\quad\text{where}\quad R_{ab}{}^{de}h_{dec'}=H_{abc'}\\
	\text{(vi$'$)}\quad&\{M_a,\hat q\Gamma_{11'}\}\xi=0\\
	\text{(vii$'$)}\quad&qM_{\hat a}\Gamma_{9'}\xi=0=q\hat qM_{\hat a}\Gamma_{9'}\Gamma_{11'}\xi\\
  \text{(viii$'$)}\quad&q=1\quad\text{\emph{or}}\quad\tilde a=\{\}\,,(\text{i.e. }a=(\hat a, a'))\\
	\text{(ix$'$)}\quad&[\hat S,\hat q\Gamma_{11'}]=0\\
	\text{(x$'$)}\quad&S=\hat S+\Gamma_{11}\hat S\Gamma_{11}\\
	\text{(xi$'$)}\quad&S\Gamma^{\hat a}(\hat S-\Gamma_{11}\hat S\Gamma_{11})=-2H^{\hat a\hat bc'}S\Gamma_{\hat bc'}\Gamma_{11}\\
	\text{(xii$'$)}\quad&S\Gamma^{\tilde a}(\hat S-\Gamma_{11}\hat S\Gamma_{11})=-R_{[\tilde a]}\varepsilon^{\tilde a\tilde b\tilde c}R_{\tilde b\tilde c}{}^{\tilde d\tilde e}\Gamma_{\tilde d\tilde e}S
\end{align*}
\caption{Group II: Conditions for backgrounds with at least one $AdS_2$ or $S^2$ factor.}
\label{table:3}
\end{table}

To conclude this section we note that the superisometry Noether current in (\ref{eq:J}) can be expanded in terms of the Killing vector, its covariant derivative and the Killing spinor as
\begin{equation}
J=J^ak_a+J^{ab}\nabla_ak_b+J_\alpha\xi^\alpha\,.
\end{equation}
Separating the conservation equation, $d*J=0$, into components correspondingly and using the Killing spinor equation and the symmetric space relations (\ref{eq:kkcom}) we find that the conservation of $J$ is equivalent to the following equations for the components
\begin{eqnarray}
&&\nabla*J^a+*J^{bc}e^dR_{bcd}{}^a=0\,,\qquad\left(\nabla*J^{ab}-*J^ae^b\right)\nabla_ak_b=0\,,\nonumber\\
&&\nabla*J_\beta\mathcal P^\beta{}_\alpha+\frac18e^a*J_\beta(M_a\mathcal P)^\beta{}_\alpha=0\,.
\label{eq:conservationids}
\end{eqnarray}
In fact we will find that, essentially due to kappa symmetry, more general versions of these equations hold. This fact will turn out to be important in the construction of the Lax connection in the following.

\subsection{Order $\Theta^0$: Bosonic Lax connection}
The bosonic terms in the conserved current take the form (\ref{eq:J01})
\begin{equation}
J^{(0)}=e^a\,k_a+*\lambda\,,
\end{equation}
where $\lambda$ satisfies $d\lambda=\frac12e^be^aH_{abc}k^c$. Since we assume $H_{abc}$ to be constant it is easy to see that we can write
\begin{equation}
\lambda=\lambda^ak_a+\lambda^{ab}\nabla_ak_b=\omega^{\hat a\hat b}h_{\hat a\hat bc'}k^{c'}-\frac12\left(e^ah_{abc}+e^{a'}h_{a'bc}\right)\nabla^bk^c\,,
\label{eq:lambda}
\end{equation}
where $h_{abc}$ is anti-symmetric and was defined in (v) and (v$'$) of table \ref{table:2} and \ref{table:3}, $\omega^{\hat a\hat b}$ id the $AdS_2$ ($S^2$) spin connection and we recall that primed indices run over the flat directions. Let us also recall the form of the string equations of motion to lowest order in $\Theta$, which is all we will need. From (\ref{eq:eom}) we have
\begin{equation}
\nabla*e^a+\frac12e^ce^bH^a{}_{bc}=\mathcal O(\Theta^2)\,,\qquad*e^a\,(\Gamma_a\mathcal D\Theta)^\alpha-e^a\,(\Gamma_a\Gamma_{11}\mathcal D\Theta)^\alpha=\mathcal O(\Theta^3)\,,
\label{eq:xeom}
\end{equation}
where $\mathcal D=\nabla+\frac18e^aM_a$. The first equation can also be derived from the first conservation equation in (\ref{eq:conservationids}) by using the form of $J^{(0)}$.

The bosonic Lax connection splits up into separate Lax connections for each factor in the geometry. Let us first consider the example of an $S^3$ factor in the geometry. We claim that the following Lax connection does the job
\begin{equation}
L^{(0)}|_{S^3}=(\alpha J^{(0)}+\beta*J^{(0)})|_{S^3}=
\alpha e^{\tilde a}k_{\tilde a}+\beta*e^{\tilde a}k_{\tilde a}
-\frac{\alpha}{2}*e^{\tilde a}h_{\tilde a\tilde b\tilde c}\,\nabla^{\tilde b}k^{\tilde c}
-\frac{\beta}{2}e^{\tilde a}h_{\tilde a\tilde b\tilde c}\,\nabla^{\tilde b}k^{\tilde c}
\,.
\end{equation}
Here $\alpha$ (not to be confused with $\alpha$ appearing in $D(2,1;\alpha)$!) and $\beta$ are parameters to be determined (they coincide with $\alpha_1$ and $\alpha_2$ of \cite{Sorokin:2010wn}). Using the fact that $\nabla e^a=0$ (i.e. the torsion of the bosonic background vanishes) and the bosonic equation of motion in (\ref{eq:xeom}) together with the symmetric space relations (\ref{eq:kkcom}) the curvature of the Lax connection becomes
\begin{eqnarray}
\lefteqn{dL^{(0)}|_{S^3}-L^{(0)}|_{S^3}L^{(0)}|_{S^3}=
\frac{\beta}{2}e^{\tilde a}e^{\tilde b}(H_{\tilde a\tilde b\tilde c}-h_{\tilde a\tilde d\tilde e}R^{\tilde d\tilde e}{}_{\tilde b\tilde c})k^{\tilde c}
}
\nonumber\\
&&{}
+\frac12((1+\alpha)\alpha-\beta^2)e^{\tilde a}*e^{\tilde b}h_{\tilde b\tilde d\tilde e}R^{\tilde d\tilde e}{}_{\tilde a\tilde c}k^{\tilde c}
+\frac12(\beta^2-\alpha^2-2\alpha)e^{\tilde a}e^{\tilde b}\nabla_{\tilde a}k_{\tilde b}
\nonumber\\
&&{}
-\frac{\alpha}{4}e^{\tilde a}e^{\tilde b}H_{\tilde a\tilde b\tilde c}h^{\tilde c\tilde d\tilde e}\,\nabla_{\tilde d}k_{\tilde e}
+\frac14(\alpha^2-\beta^2)e^{\tilde a}e^{\tilde b}h_{\tilde b\tilde c\tilde d}h_{\tilde a\tilde e\tilde f}R^{\tilde e\tilde f\tilde c}{}_{\tilde g}\nabla^dk^g\,.
\end{eqnarray}
Using condition (v) of table \ref{table:2} the right-hand-side reduces to
\begin{eqnarray}
\frac{1-q^2}{2}(\beta^2-\alpha^2-2\alpha)e^{\tilde a}e^{\tilde b}\nabla_{\tilde a}k_{\tilde b}
\end{eqnarray}
which vanishes provided that\footnote{For the special case $q=1$ it appears we don't need to impose any condition, however this condition will arise again at higher orders in $\Theta$.}
\begin{equation}
\beta^2=\alpha^2+2\alpha\,.
\label{eq:spectral}
\end{equation}
Since there is only one condition on two parameters we have one free (spectral) parameter and a one-parameter family of flat connections.

Let us now consider the Lax connection for an $S^2$-factor in the geometry. We will find that we need to modify two of the coefficients in the Lax connection and we will take it to have the form
\begin{equation}
L^{(0)}|_{S^2}=
\hat\alpha e^{\hat a}k_{\hat a}+\hat\beta*e^{\hat a}k_{\hat a}
-\frac{\alpha}{2}*e^{a'}h_{a'\hat b\hat c}\,\nabla^{\hat b}k^{\hat c}
-\frac{\beta}{2}e^{a'}h_{a'\hat b\hat c}\,\nabla^{\hat b}k^{\hat c}
\,,
\end{equation}
where $\alpha$ and $\beta$ are the same as before satisfying (\ref{eq:spectral}) and $\hat\alpha$ and $\hat\beta$ are to be determined. Computing the curvature in the same way as before we find, keeping in mind condition (iv$'$) in table \ref{table:3},
\begin{eqnarray}
\lefteqn{dL^{(0)}|_{S^2}-L^{(0)}|_{S^2}L^{(0)}|_{S^2}
=
\frac12(\alpha(1+\hat\alpha)-\beta\hat\beta)e^{\hat a}*e^{b'}h_{b'\hat c\hat d}R^{\hat c\hat d}{}_{\hat a\hat e}k^{\hat e}
}
\nonumber\\
&&{}
-\frac12(\alpha\hat\beta-(1+\hat\alpha)\beta)e^{\hat a}e^{b'}h_{b'\hat c\hat d}R^{\hat c\hat d}{}_{\hat a\hat e}k^{\hat e}
-\hat\beta e^{\hat a}e^{b'}H_{b'\hat a\hat c}k^{\hat c}
+\frac12(\hat\beta^2-\hat\alpha^2-2\hat\alpha)e^{\hat a}e^{\hat b}\nabla_{\hat a}k_{\hat b}
\nonumber\\
&&{}
-\frac{\alpha}{4}e^{\hat a}e^{\hat b}H_{\hat a\hat bc'}h^{c'\hat d\hat e}\,\nabla_{\hat d}k_{\hat e}
+\frac14(\beta^2-\alpha^2)e^{a'}e^{b'}h_{a'\hat c\hat d}h_{b'\hat e\hat f}R^{\hat e\hat f\hat c}{}_{\hat g}\nabla^{\hat d}k^{\hat g}\,.
\end{eqnarray}
Using condition (v$'$) the right-hand-side now reduces to
\begin{eqnarray}
&&\frac12(\alpha(1+\hat\alpha)-\beta\hat\beta)e^{\hat a}*e^{b'}H_{b'\hat a\hat c}k^{\hat c}
-\frac12((2+\alpha)\hat\beta-(1+\hat\alpha)\beta)e^{\hat a}e^{b'}H_{b'\hat a\hat c}k^{\hat c}
\nonumber\\
&&{}
+\frac12(\hat\beta^2-\hat\alpha^2-2\hat\alpha+\frac12q^2\alpha)e^{\hat a}e^{\hat b}\nabla_{\hat a}k_{\hat b}\,.
\end{eqnarray}
We now observe that these terms vanish if we take
\begin{equation}
\hat\alpha=\gamma(2+\alpha)-1\,,\qquad\hat\beta=\gamma\beta\qquad\text{where}\qquad\gamma=\frac12\sqrt{\frac{2+q^2\alpha}{2+\alpha}}\,,
\label{eq:spectralhat}
\end{equation}
giving us again a one-parameter family of flat connections.

Putting together what we have learned we write the total Lax connection
\begin{equation}
L^{(0)}=
\alpha_{[a]}e^a\tilde k_a
+\beta_{[a]}(*e^a+\lambda^a)\tilde k_a
-\frac{\alpha}{2}*e^ah_{abc}\nabla^b\tilde k^c
-\frac{\beta}{2}e^ah_{abc}\nabla^b\tilde k^c\,,
\label{eq:L0}
\end{equation}
where 
\begin{equation}
\alpha_{[\hat a]}=\hat\alpha\,,\qquad\beta_{[\hat a]}=\hat\beta\,,\qquad\alpha_{[\tilde a\,,a']}=\alpha\,,\qquad\beta_{[\tilde a\,,a']}=\beta\,.
\label{eq:alphaa}
\end{equation}
Two remarks should be made about the form of $L^{(0)}$ in (\ref{eq:L0}). Firstly we have included a factor of $\lambda^a$ together with $*e^a$ in the second term. This only affects the term involving $k_{a'}$ since $\lambda^a=\lambda^{a'}$ by (\ref{eq:lambda}) and it is in fact needed in that term so that there are no $k_{a'}$ terms generated when computing $d$ of $L^{(0)}$. Note that 
\begin{equation}
*J^{(0)\,a}=*e^a+\lambda^a\,,\qquad *J^{(0)\,ab}=-\frac{c_{[a]}}{2}e^ch_c{}^{ab}\,,
\end{equation}
with $c_{[a]}$ defined in (\ref{eq:ca}), so that the Lax connection is indeed built from the components of the conserved current as in \cite{Sorokin:2010wn}. The next important comment is that we have written $\tilde k_a$ instead of $k_a$. The reason for this will become clear when we try to find $L^{(2)}$. The $\tilde k_a$ still satisfy the same algebraic relations as those written for $k_a$ in the previous section however. The reason for the notation is that we will find that in the examples in group II with an $AdS_3$ ($S^3$) factor only the left $SL(2,\mathbbm R)_L$ ($SU(2)_L$) isometry generators should appear in $L^{(0)}$ (these are the ones that sit inside $D(2,1;\alpha)$, see table \ref{table:1}).

\subsection{Order $\Theta^1$ Lax connection}
We will now show that the terms in the Lax connection linear in $\Theta$ take the form
\begin{equation}
L^{(1)}=*J^{(1)}_\alpha(VW\xi)^\alpha\,,
\label{eq:L1}
\end{equation}
where $V$ and $W$ are matrices (that depend on the spectral parameter) to be determined. The superisometry Noether current at linear order in $\Theta$ is $J^{(1)}=J^{(1)}_\alpha\xi^\alpha$ and from (\ref{eq:J01}) we read off that
\begin{equation}
J^{(1)}_\alpha=ie^a\,(\Theta\Gamma_a)_\alpha-i*e^a\,(\Theta\Gamma_a\Gamma_{11})_\alpha\,.
\label{eq:J1alpha}
\end{equation}
Note that we do not impose the projection by $\mathcal P$ on $J^{(1)}_\alpha$ (recall that $\xi=\mathcal P\xi$). This will be important as the matrices $V\,,W$ will typically not commute with $\mathcal P$. Even without the projection by $\mathcal P$ an equation like the conservation equation in (\ref{eq:conservationids}) is still satisfied. Indeed we have, using the form of $J^{(1)}_\alpha$ and the lowest order equations of motion (\ref{eq:xeom})
\begin{eqnarray}
\nabla*J^{(1)}_\alpha
&=&
i\nabla*e^a\,(\Theta\Gamma_a)_\alpha
+\frac{i}{8}*e^ae^b\,(\Theta[-H_{bcd}\Gamma^{cd}\Gamma_{11}+\Gamma_bS]\Gamma_a)_\alpha
\nonumber\\
&&{}
-\frac{i}{8}e^ae^b\,(\Theta[-H_{bcd}\Gamma^{cd}\Gamma_{11}+\Gamma_bS]\Gamma_a\Gamma_{11})_\alpha
+\mathcal O(\Theta^3)
\nonumber\\
&=&
-\frac18e^a*J^{(1)}_\beta(M_a)^\beta{}_\alpha
+\mathcal O(\Theta^3)\,,
\end{eqnarray}
where we used the fact that $M_a=H_{abc}\Gamma^{bc}\Gamma_{11}+S\Gamma_a$.

Using the form of $L^{(0)}$ in (\ref{eq:L0}) and $L^{(1)}$ in (\ref{eq:L1}), the commutation relation (\ref{eq:kxi-comm}), and noting that $*J^{(1)}_\alpha=-(J^{(1)}\Gamma_{11})_\alpha$ we find the curvature of the Lax connection at linear order in $\Theta$ to be
\begin{eqnarray}
\lefteqn{d(L^{(0)}+L^{(1)})-(L^{(0)}+L^{(1)})(L^{(0)}+L^{(1)})=
dL^{(1)}
-L^{(0)}L^{(1)}
-L^{(1)}L^{(0)}
+\mathcal O(\Theta^2)}
\nonumber\\
&=&
\frac18e^aJ^{(1)}_\alpha(\Gamma_{11}M_aVW\xi)^\alpha
-\frac18e^aJ^{(1)}_\alpha(\Gamma_{11}[1+\alpha_{[a]}+\beta_{[a]}\Gamma_{11}]VWM_a\xi)^\alpha
\nonumber\\
&&{}
+\frac18e^ah_{abc}R_{de}{}^{bc}J^{(1)}_\alpha([\alpha+\beta\Gamma_{11}]VW\Gamma^{de}\xi)^\alpha
-\frac18\beta_{[a]}\lambda^aJ^{(1)}_\alpha(\Gamma_{11}VWM_a\xi)^\alpha
+\mathcal O(\Theta^2)\,.
\end{eqnarray}
The last term vanishes due to (iii) since $\lambda^a=\lambda^{a'}$. Taking the matrix $V$ to have the form
\begin{equation}
V=a-b\Gamma_{11}\,,\qquad\mbox{where}\qquad a=\frac{\beta}{\sqrt{2\alpha}}\,,\qquad b=\sqrt{\frac{\alpha}{2}}
\label{eq:V}
\end{equation}
it is easy to see, using (\ref{eq:spectral}) and (\ref{eq:spectralhat}), that $V$ satisfies the relations
\begin{equation}
VV^\dagger=a^2-b^2=1\,,\qquad V^2=1+\alpha-\beta\Gamma_{11}\,,\qquad(1+\hat\alpha+\hat\beta\Gamma_{11})V=\gamma(2+\alpha+\beta\Gamma_{11})V=2a\gamma\,.
\end{equation}
The curvature of the Lax connection reduces to
\begin{eqnarray}
&&\frac18e^aJ^{(1)}_\alpha(\Gamma_{11}M_aVW\xi)^\alpha
-\frac{a\gamma}{4}e^{\hat a}J^{(1)}_\alpha(\Gamma_{11}WM_{\hat a}\xi)^\alpha
-\frac18e^{\tilde a}J^{(1)}_\alpha(V^\dagger\Gamma_{11}WM_{\tilde a}\xi)^\alpha
\nonumber\\
&&{}
+\frac{b}{4}e^{\tilde a}H_{\tilde a\tilde b\tilde c}J^{(1)}_\alpha(\Gamma_{11}W\Gamma^{\tilde b\tilde c}\xi)^\alpha
+\frac{b}{4}e^{a'}H_{a'\hat b\hat c}J^{(1)}_\alpha(\Gamma_{11}W\Gamma^{\hat b\hat c}\xi)^\alpha\,,
\end{eqnarray}
where we have used (iv), (v), (iv$'$) and (v$'$) to rewrite the term with $h$. Using the form of $M_a$ in (\ref{eq:Killing}) and assuming that $W$ commutes with $H_{a'\hat b\hat c}\Gamma^{\hat b\hat c}$ and $H_{\tilde a\tilde b\tilde c}\Gamma^{\tilde b\tilde c}$ we get
\begin{eqnarray}
&&\frac18e^{\hat a}J^{(1)}_\alpha(\Gamma_{11}M_{\hat a}VW\xi)^\alpha
-\frac{a\gamma}{4}e^{\hat a}J^{(1)}_\alpha(\Gamma_{11}WM_{\hat a}\xi)^\alpha
+\frac18e^{\tilde a}J^{(1)}_\alpha(V^\dagger\Gamma_{11}[M_{\tilde a},W]\xi)^\alpha
\nonumber\\
&&{}
+\frac18e^{a'}J^{(1)}_\alpha(V^\dagger\Gamma_{11}M_{a'}W\xi)^\alpha\,.
\end{eqnarray}
It is clear that due to (vi) and (iii) this vanishes for Group I if we take $W=c+qd\Gamma^{9'}$ (note that in this case the first two terms are absent). The free coefficients will be fixed at the next order and we will get
\begin{equation}
W_{\text{I}}=\frac{1}{2\sqrt2}(\beta+q\alpha\Gamma^{9'})\qquad\Rightarrow\qquad W_{\text{I}}W_{\text{I}}^\dagger=\frac18(2+\hat q^2\alpha)\alpha
%\,,\qquad W_{\text{I}}^2=\frac18(\beta^2+q^2\alpha^2+2q\alpha\beta\Gamma^{9'})
\,.
\label{eq:WI}
\end{equation}
For Group II consider taking $W=c+\hat qd\Gamma_{11'}$. The third term vanishes due to (viii$'$) and the last term vanishes due to (vi$'$) and (iii). Using the form of $V$, (vi$'$), (vii$'$) and the fact that $\Gamma_{11}=\hat q\Gamma_{11'}-q\Gamma_{9'}$ we get
\begin{eqnarray}
\frac{b}{8\alpha}e^{\hat a}J^{(1)}_\alpha(\Gamma_{11}[(\beta+\hat q\alpha\Gamma_{11'})W^\dagger-2\gamma\beta W]M_{\hat a}\xi)^\alpha\,.
\end{eqnarray}
Which vanishes for the following choice of $W$ (again the overall normalization is determined at the next order in $\Theta$)
\begin{equation}
%VX=c(a+b\Gamma_{11'})[2\gamma(2+\alpha)+q\beta\Gamma_{9'}]
%a^2=1/(2\gamma)+1
%b^2=1/(2\gamma)-1
%ab=-\hat q\alpha/(2\gamma\beta)
%c=\gamma\sqrt\alpha/(2\sqrt2)
W_{\text{II}}=\frac{1}{2\sqrt2}\sqrt{\frac{\gamma}{1+2\gamma}}((1+2\gamma)\beta+\hat q\alpha\Gamma_{11'})\quad\Rightarrow\quad W_{\text{II}}W_{\text{II}}^\dagger=\frac12\hat\beta^2\,,\quad
W_{\text{II}}^2=\frac{\hat\beta}{4}(\beta+\hat q\alpha\Gamma_{11'})\,,
\label{eq:WII}
\end{equation}
as is easily seen using (\ref{eq:spectral}), (\ref{eq:spectralhat}) and the fact that $q^2+\hat q^2=1$. This concludes the demonstration of the flatness of $L^{(0)}+L^{(1)}$ modulo terms of order $\Theta^2$.

Let us end this section by computing $L^{(1)}L^{(1)}$ which we will need for our analysis of the $\Theta^2$-terms in the next section. Using the form of $L^{(1)}$ in (\ref{eq:L1}) and that of $J^{(1)}_\alpha$ in (\ref{eq:J1alpha}) we get
\begin{align}
L^{(1)}L^{(1)}&=
-\frac12e^ae^b\,\Theta\Gamma_aVW\{\xi,\xi\}W^\dagger V^\dagger\Gamma_b\Theta
-\frac12e^ae^b\,\Theta\Gamma_a\Gamma_{11}VW\{\xi,\xi\}W^\dagger V^\dagger\Gamma_{11}\Gamma_b\Theta
\label{eq:L1L1prime}
\\
&\qquad
+\frac12*e^ae^b\,\Theta\Gamma_a\Gamma_{11}VW\{\xi,\xi\}W^\dagger V^\dagger\Gamma_b\Theta
+\frac12*e^ae^b\,\Theta\Gamma_aVW\{\xi,\xi\}W^\dagger V^\dagger\Gamma_{11}\Gamma_b\Theta\,.
\nonumber
\end{align}
Using the form of the commutator of two Killing spinors in (\ref{eq:xixi-comm}), the constraints in table \ref{table:2} and \ref{table:3}, the form of $V$ in (\ref{eq:V}) and the form of $W$ in (\ref{eq:WI}) and (\ref{eq:WII}) one gets after some algebra (the details of the calculation have been deferred to appendix \ref{sec:L1L1})
\begin{eqnarray}
L^{(1)}L^{(1)}&=&
\frac{i}{32}\alpha\beta
\left(
e^ce^d\,\Theta\Gamma_c\Gamma^{\tilde a\tilde b}S\Gamma_d\Theta
-*e^ce^d\,\Theta\Gamma_c\Gamma^{\tilde a\tilde b}S\Gamma_d\Theta
\right)
H_{\tilde a\tilde b\tilde e}k^{\tilde e}
\nonumber\\
&&{}
+\frac{i}{32}\hat\beta
\left(
e^ce^d\,\Theta\Gamma_c\Gamma^{\hat ab'}(\alpha-\beta\Gamma_{11})S\Gamma_d\Theta
-*e^ce^d\,\Theta\Gamma_c\Gamma^{\hat ab'}(\alpha-\beta\Gamma_{11})\Gamma_{11}S\Gamma_d\Theta
\right)
H_{\hat ab'\hat e}k^{\hat e}
\nonumber\\
&&{}
+\frac{i}{32}\beta^2_{[a]}
\left(
e^ce^d\,\Theta\Gamma_c\Gamma^{ab}S\Gamma_d\Theta
-*e^ce^d\,\Theta\Gamma_c\Gamma^{ab}\Gamma_{11}S\Gamma_d\Theta
\right)\widetilde{\nabla_ak_b}
\nonumber\\
&&{}
-\frac{i}{64}\alpha^2
\left(
e^ce^d\,\Theta\Gamma_c\Gamma^{\tilde a\tilde b}S\Gamma_d\Theta
-*e^ce^d\,\Theta\Gamma_c\Gamma^{\tilde a\tilde b}\Gamma_{11}S\Gamma_d\Theta
\right)h_{\tilde a\tilde f\tilde g}H^{\tilde g}{}_{\tilde e\tilde b}\nabla^{\tilde e}k^{\tilde f}\,.
\label{eq:L1L1}
\end{eqnarray}
Here we have defined $\widetilde{\nabla_ak_b}$ which is the same as $\nabla_ak_b$ \emph{except} for group II where
\begin{eqnarray}
\widetilde{\nabla_{\tilde a}k_{\tilde b}}=
\frac12\Big(\nabla_{\tilde a}k_{\tilde b}+\frac{R_{[\tilde e]}}{2}R_{\tilde a\tilde b\tilde c\tilde d}\varepsilon^{\tilde c\tilde d\tilde e}k_{\tilde e}\Big)\,.
\label{eq:tildenablak}
\end{eqnarray}

\subsection{Order $\Theta^2$ Lax connection}
Computing the order $\Theta^2$ terms in the curvature of $L^{(0)}+L^{(1)}$ using the superisometry algebra of section \ref{sec:algebra}, the conservation of $J$ in (\ref{eq:conservationids}) as well as the lowest order equations of motion (\ref{eq:xeom}) we find
\begin{eqnarray}
F^{(2)}&=&
-d\left(
\beta_{[a]}*J^{(2)\,a}\tilde k_a
+\beta_{[a]}*J^{(2)\,ab}\nabla_a\tilde k_b
-\frac{\alpha}{2}*J^{(2)}_ah^{abc}\,\nabla_b\tilde k_c
\right)
\nonumber\\
&&{}
-\frac{\alpha}{2}*e^aJ^{(2)}_bh^{bcd}R_{cdae}\tilde k^e
+\frac{\alpha}{2}*e^aJ^{(2)\,bc}R_{bcad}h^{def}\nabla_e\tilde k_f
-L^{(1)}L^{(1)}
\label{eq:F2}
\,.
\end{eqnarray}
Recall from the discussion in section \ref{sec:deformation} that we need to show that this can be written as $DL^{(2)}=dL^{(2)}-L^{(0)}L^{(2)}-L^{(2)}L^{(0)}$, for some $L^{(2)}$, for the Lax connection to exist up to quadratic order in $\Theta$. To do this we will need two identities. Using the lowest order bosonic and fermionic equations of motion in (\ref{eq:xeom}) and the form of the conserved current in (\ref{eq:J2}) we have 
\begin{eqnarray}
\nabla*J^{(2)\,ab}+*e^{[a}J^{(2)\,b]}
&=&
-\frac12e^{[a}J^{(2)}_{cd}H^{b]cd}
-e^cJ^{(2)\,d[a}H^{b]}{}_{cd}
+\frac{i}{32}*e^ce^d\,\Theta\Gamma_c\Gamma^{ab}S\Gamma_d\Theta
\nonumber\\
&&{}
-\frac{i}{32}e^ce^d\,\Theta\Gamma_c\Gamma^{ab}\Gamma_{11}S\Gamma_d\Theta
+\mathcal O(\Theta^4)\,.
\end{eqnarray}
This is a generalized version of the second conservation equation in (\ref{eq:conservationids}). Note that the right-hand-side vanishes when contracted with $\nabla_ak_b$ as needed for consistency with the conservation equation. It is also useful to note that the first two terms on the right-hand-side vanish when contracted with $H_{abe}$ as follows from (v) and (v$'$) in table \ref{table:2} and \ref{table:3}. We can also derive a ''dual'' version of this equation. Using the form of $J^{(2)}$, the conservation of $J$ in (\ref{eq:conservationids}) and the lowest order equations of motion in (\ref{eq:xeom}) we find
\begin{eqnarray}
\nabla J^{(2)\,ab}+e^{[a}J^{(2)\,b]}
&=&
-\frac12*e^{[a}J^{(2)}_{cd}H^{b]cd}
+*e^cJ^{(2)d[a}H^{b]}{}_{cd}
+\frac{i}{32}e^ce^d\,\Theta\Gamma_c\Gamma^{ab}S\Gamma_d\Theta
\nonumber\\
&&{}
-\frac{i}{32}*e^ce^d\,\Theta\Gamma_c\Gamma^{ab}\Gamma_{11}S\Gamma_d\Theta
+\mathcal O(\Theta^4)\,.
\end{eqnarray}
Using these two equations, the form of $L^{(0)}$ in (\ref{eq:L0}), the superisometry algebra in section \ref{sec:algebra}, the conditions (iv), (v), (iv$'$) and (v$'$) in table \ref{table:2} and \ref{table:3} and the expression for $L^{(1)}L^{(1)}$ given in (\ref{eq:L1L1}) one can show, after a bit of algebra, that
\begin{eqnarray}
&&D\Big(
\beta_{[a]}*J^{(2)\,a}\tilde k_a
-\frac{\alpha}{2}*J^{(2)}_ah^{abc}\,\nabla_b\tilde k_c
+\beta_{[a]}\big[(1+\alpha_{[a]})*J^{(2)\,ab}+\beta_{[a]}J^{(2)\,ab}\big]\nabla_a\tilde k_b
\nonumber\\
&&{}
+\alpha\big[\Big(\frac32+\alpha\Big)*J^{(2)\tilde a\tilde b}+\beta J^{(2)\,\tilde a\tilde b}\big]H_{\tilde a\tilde b\tilde c}\tilde k^{\tilde c}
+\hat\beta\big[\beta*J^{(2)\,a'\hat b}+\alpha J^{(2)\,a'\hat b}\big]H_{a'\hat b\hat c}\tilde k^{\hat c}
\nonumber\\
&&{}
-\frac{\alpha}{2}\big[\beta*J^{(2)\,\tilde a\tilde b}+\alpha J^{(2)\,\tilde a\tilde b}\big]h_{\tilde b\tilde c}{}^{\tilde e}H_{\tilde e\tilde d\tilde a}\nabla^{\tilde c}\tilde k^{\tilde d}
\Big)
\nonumber\\
&=&
-F^{(2)}
+\frac{i}{64}(3+2\alpha)\alpha
\left(
*e^ce^d\,\Theta\Gamma_c\Gamma^{\tilde a\tilde b}S\Gamma_d\Theta
-e^ce^d\,\Theta\Gamma_c\Gamma^{\tilde a\tilde b}\Gamma_{11}S\Gamma_d\Theta
\right)
H_{\tilde a\tilde b\tilde e}\tilde k^{\tilde e}
\nonumber\\
&&{}
-\frac{i}{64}\alpha\beta
\left(
*e^ce^d\,\Theta\Gamma_c\Gamma^{\tilde a\tilde b}S\Gamma_d\Theta
-e^ce^d\,\Theta\Gamma_c\Gamma^{\tilde a\tilde b}\Gamma_{11}S\Gamma_d\Theta
\right)
h_{\tilde a\tilde f}{}^{\tilde g}H_{\tilde g\tilde e\tilde b}\nabla^{\tilde e}\tilde k^{\tilde f}
\nonumber\\
&&{}
-\frac{\alpha}{2}e^{\tilde a}*J^{(2)\,\tilde b\tilde c}R_{\tilde b\tilde c\tilde a\tilde e}h^{\tilde e\tilde f\tilde d}\nabla_{\tilde f}\tilde k_{\tilde d}
-\frac{\alpha}{2}e^{\tilde a}*J^{(2)\,\tilde b\tilde c}H_{\tilde b\tilde c\tilde d}\nabla_{\tilde a}\tilde k^{\tilde d}
+\frac32\alpha\beta e^{[\tilde a}J^{(2)\,\tilde b\tilde c]}H_{\tilde a\tilde b\tilde d}\nabla_{\tilde c}\tilde k^{\tilde d}
\nonumber\\
&&{}
+\frac32q^2\alpha^2e^{[\tilde a}*J^{(2)\,\tilde b\tilde c]}H_{\tilde a\tilde b\tilde d}\nabla_{\tilde c}\tilde k_{\tilde d}
+\hat\beta^2e^{a'}*J^{(2)\,\hat b\hat c}H_{a'\hat b\hat d}\nabla_{\hat c}\tilde k_{\hat d}
-2\hat\beta^2e^{[\hat a}*J^{(2)\,\hat b]c'}H_{c'\hat a\hat d}\nabla_{\hat b}\tilde k^{\hat d}\,,
\nonumber\\
\label{eq:DL2}
\end{eqnarray}
where $F^{(2)}$ was given in (\ref{eq:F2}). Here we have had to assume a specific form of $\tilde k_a$. For group I $\tilde k_a=k_a$ while for group II $\tilde k_{\hat a,a'}=k_{\hat a,a'}$ while
\begin{equation}
\tilde k_{\tilde a}=\frac12\Big(k_{\tilde a}-\frac{R_{[\tilde a]}}{2}\varepsilon_{\tilde a\tilde b\tilde c}\nabla^{\tilde b}k^{\tilde c}\Big)\,.
\label{eq:tildek}
\end{equation}
This form of $\tilde k_{\tilde a}$ is chosen such that $\nabla_{\tilde a}\tilde k_{\tilde b}=\widetilde{\nabla_{\tilde a}k_{\tilde b}}$ defined in (\ref{eq:tildenablak}) and is needed in order to cancel the corresponding term appearing in $L^{(1)}L^{(1)}$ eq. (\ref{eq:L1L1}). It is not hard to show that $\tilde k_a$ satisfies the same algebra as $k_a$ but with the extra constraint
\begin{equation}
%[\tilde k_a,\tilde k_b]=
\nabla_{\tilde a}\tilde k_{\tilde b}=\frac{R_{[\tilde a]}}{2}R_{\tilde a\tilde b}{}^{\tilde c\tilde d}\varepsilon_{\tilde c\tilde d\tilde e}\tilde k^{\tilde e}\,.
\end{equation}
$\tilde k_a$ can be interpreted as the $SL(2,\mathbbm R)_L$ ($SU(2)_L$) isometries of $AdS_3$ ($S^3$). The reason only the left isometries appear is because these are the ones sitting inside $D(2,1;\alpha)$ and hence the only ones that can appear in the commutator of two Killing spinors and therefore in $L^{(1)}L^{(1)}$.

In fact all terms on the right-hand-side of (\ref{eq:DL2}) except for $F^{(2)}$ vanish! The terms involving $S$ vanish by the symmetry properties of the gamma-matrices and the fact that $S$ commutes with the relevant combinations of gamma matrices. The next two terms cancel upon using the form of $R_{ab}{}^{cd}\sim\delta_{[a}^c\delta_{b]}^d$ and (v) of table \ref{table:2} while the rest of the terms vanish due to the anti-symmetry in the indices (note that in these terms the tilded indices can only run over three values while the hatted ones can only take two values). We have therefore managed to write $F^{(2)}$ as $D$ of something proving that the Lax connection can be extended to the quadratic order in $\Theta$.

\subsection{Summary}\label{sec:summary}
We have shown that the string in a background with constant fluxes satisfying the conditions listed in section \ref{sec:algebra} is classically integrable by constructing its Lax connection up to quadratic order in $\Theta$. The Lax connection takes the form
\begin{equation}
L=L^{(0)}+L^{(1)}+L^{(2)}+\mathcal O(\Theta^3)\,,
\end{equation}
where the different pieces are built from components of the superisometry current as
\begin{align}
L^{(0)}&=
\alpha_{[a]}e^a\tilde k_a
+\beta_{[a]}*J^{(0)\,a}\tilde k_a
-\frac{\alpha}{2}*e^ah_{abc}\nabla^bk^c
-\frac{\beta}{2}e^ah_{abc}\nabla^bk^c\,,\qquad
L^{(1)}=*J^{(1)}_\alpha(VW\xi)^\alpha\,,
\nonumber\\
&\nonumber\\
L^{(2)}&=
\beta_{[a]}*J^{(2)\,a}\tilde k_a
-\frac{\alpha}{2}*J^{(2)}_ah^{abc}\nabla_bk_c
+\beta_{[a]}\big[(1+\alpha_{[a]})*J^{(2)\,ab}+\beta_{[a]}J^{(2)\,ab}\big]\nabla_a\tilde k_b
\nonumber\\
&\qquad
+\alpha\big[\big(\frac32+\alpha\big)*J^{(2)\tilde a\tilde b}+\beta J^{(2)\,\tilde a\tilde b}\big]H_{\tilde a\tilde b\tilde c}k^{\tilde c}
+\hat\beta\big[\beta*J^{(2)\,a'\hat b}+\alpha J^{(2)\,a'\hat b}\big]H_{a'\hat b\hat c}k^{\hat c}
\nonumber\\
&\qquad
-\frac{\alpha}{2}\big[\beta*J^{(2)\,\tilde a\tilde b}+\alpha J^{(2)\,\tilde a\tilde b}\big]h_{\tilde b\tilde c}{}^{\tilde e}H_{\tilde e\tilde d\tilde a}\nabla^{\tilde c}k^{\tilde d}\,,
\end{align}
where $V$ and $W$ are defined in (\ref{eq:V}), (\ref{eq:WI}) and (\ref{eq:WII}), $\tilde k_a$ defined in (\ref{eq:tildek}) and $\alpha_{[a]}$ and $\beta_{[a]}$ in (\ref{eq:alphaa}).

Note that for the construction of the Lax connection at quadratic order, $L^{(2)}$, it was important that we took the most general form of $L^{(0)}$ and $L^{(1)}$. For example, taking $L^{(0)}$ with $k_a$ instead of $\tilde k_a$ is perfectly fine up to order $\Theta^1$ but we would fail to construct $L^{(2)}$ since only terms involving the left isometries $SL(2,\mathbbm R)_L$ ($SU(2)_L$) are generated from $L^{(1)}L^{(1)}$ as we have seen. Similarly it would have been fine to drop the $\Gamma^{9'}$-term in $W_{\text{I}}$ at order $\Theta^1$ but this term is needed in order to construct $L^{(2)}$ since it affects the terms coming from $L^{(1)}L^{(1)}$.

Finally, let us imagine that we changed some signs of the fluxes in the supergravity solution so as to break the supersymmetry. The bosonic Lax connection would still be fine but there would be no $L^{(1)}$ since there would no longer be any Killing spinors from which to construct it.\footnote{One could try instead to construct an $L^{(1)}$ using some constant spinor or the Killing spinor of the supersymmetric background. However, even if this worked to linear order in $\Theta$ it would not fix the problem at quadratic order since $L^{(1)}L^{(1)}$ would not be of the required form.} Trying to go to the next order one would then encounter a problem. The terms in the curvature that were canceled by $L^{(1)}L^{(1)}$ in the supersymmetric case (\ref{eq:L1L1}) would now be left over. They would certainly not vanish in general and it does not appear possible to write them as $D$ of something. This argument suggests that kappa symmetry of the string alone is not enough to ensure that integrability of the bosonic string lifts to the superstring. A very interesting question is whether kappa symmetry together with some amount of supersymmetry is enough for the bosonic integrability to extend to the full superstring. We hope to return to this question in the future.

\section{Integrable type IIA backgrounds from intersecting branes}\label{sec:backgrounds}
In this section we will give the details of the backgrounds listed in table \ref{table:1} and show that they fulfill all the conditions we had to impose in the previous section to prove the integrability. Our conventions for the volume form and curvature of $AdS_n$ and $S^n$ are
\begin{align}
\Omega_{AdS_n}&=\frac{1}{n!}e^{a_{n-1}}\cdots e^{a_0}\varepsilon_{a_0\cdots a_{n-1}}=-e^{n-1}\cdots e^0\,,& R_{ab}{}^{cd}(AdS_n)&=\frac{2}{R^2_{AdS}}\delta_{[a}^c\delta_{b]}^d\\
\Omega_{S^n}&=\frac{1}{n!}e^{a_n}\cdots e^{a_1}\varepsilon_{a_1\cdots a_n}=e^n\cdots e^1\,,& R_{ab}{}^{cd}(S^n)&=-\frac{2}{R^2_S}\delta_{[a}^c\delta_{b]}^d\,,
\end{align}
where $R$ is the radius of curvature.

The supersymmetry conditions are (see section \ref{sec:thetaexp})
\begin{equation}
\text{Dilatino eq: }\quad T\xi=0\qquad\text{Gravitino eq: }\quad U_{ab}\xi=0\,,
\label{eq:SUSYcond}
\end{equation}
where, for constant dilaton and fluxes,
\begin{equation}
T=\frac{i}{24}H_{abc}\,\Gamma^{abc}\Gamma_{11}+\frac{i}{16}\Gamma_aS\Gamma^a\,,
\qquad
U_{ab}=\frac{1}{32}M_{[a}M_{b]}-\frac{1}{4}R_{abcd}\Gamma^{cd}\,.
\label{eq:TandUab}
\end{equation}
The Killing spinor satisfies
\begin{equation}
\xi=\mathcal P\xi\,,
\label{eq:Killing-proj}
\end{equation}
where $\mathcal P$ is a projector which projects on the supersymmetric directions in spinor space. We also introduce a parameter $0\leq q\leq1$ such that
\begin{equation}
q=0\Leftrightarrow\text{ No NSNS-flux.}
\end{equation}
Backgrounds with non-zero NSNS-flux which can not be tuned we take to have $q=1$.

Note that the specific form of the fluxes we give can be changed by performing T-dualities. This does not however affect the rest of the discussion, in particular the integrability goes through in the same way (as it must).

\subsection*{A. $AdS_4\times\mathbbm{CP}^3$}
This supergravity solution can be obtained by dimensional reduction from the maximally supersymmetric $AdS_4\times S^7$ solution arising as the near-horizon geometry of the M2-brane. This is done by viewing $S^7$ as an $S^1$ Hopf fibration over $\mathbbm{CP}^3$ and reducing on the $S^1$ \cite{Nilsson:1984bj,Sorokin:1985ap}. This breaks the superisometry group of the eleven-dimensional solution, $OSp(8|4)$, down to $OSp(6|4)$, leaving 24 unbroken supersymmetries. The type IIA solution has RR four-form and two-form flux, the latter arising through the Hopf reduction procedure. The fluxes take the form
\begin{equation}
F^{(2)}=e^{-\phi}J\,,\qquad F^{(4)}=-3e^{-\phi}\Omega_{AdS_4}\,.
\end{equation}
Here $J$ is the K\"ahler form on $\mathbbm{CP}^3$ and we take $J=e^5e^4+e^7e^6+e^9e^8$ in terms of the $\mathbbm{CP}^3$ vielbeins. Note that we have indicated the dilaton dependence even though the dilaton is just a constant for the backgrounds we consider here. From the definition of $S$ in (\ref{eq:S}) and $T$ in (\ref{eq:TandUab}) we find
\begin{equation}
S=4\mathcal P\Gamma^{0123}\,,\qquad T=\frac{3i}{2}\Gamma^{0123}(1-\mathcal P)\,,
%S=e^\phi\big(\frac12F_{ab}^{(2)}\Gamma^{ab}\Gamma_{11}+\frac{1}{4!}F_{abcd}^{(4)}\Gamma^{abcd}\big)
%2S\Gamma_{[0}S\Gamma_{1]}\xi=32\Gamma_{01}\xi
%2S\Gamma_{[4}S\Gamma_{5]}\xi=-16[\Gamma_45+(\Gamma^{45}+\Gamma^{67}+\Gamma^{89})]\xi
\end{equation}
where the Killings spinor projection matrix is given by
\begin{equation}
\mathcal P=\frac14(3+\Gamma^{6789}+\Gamma^{4589}+\Gamma^{4567})\,.
%(\Gamma^{6789}+\Gamma^{4589}+\Gamma^{4567})^2=3-2(\Gamma^{4567}+\Gamma^{4589}+\Gamma^{6789})
%eigenvalues 24x(1), 8x(-3)
\end{equation}
The dilatino equation in (\ref{eq:SUSYcond}) is obviously satisfied and it is also easy to check the gravitino equation. One finds the radii of curvature to be
\begin{equation}
R_{AdS_4}=1\,,\qquad R_{\mathbbm{CP}^3}=2\,.
\end{equation}

Since there is no NSNS flux ($q=0$) eq. (\ref{eq:kxi-comm}) is trivially true (with $\hat S=S$) while conditions (i), (ii) and (iii) are clearly true and the conditions in table \ref{table:2} become trivial.

\subsection*{B. $AdS_3\times S^3\times S^3\times S^1$ (and $AdS_3\times S^3\times T^4$)}
This supergravity background arises as the dimensional reduction of the eleven-dimensional $AdS_3\times S^3\times S^3\times T^2$ solution representing the near-horizon geometry of two M5-branes and an M2-brane intersecting over a line \cite{Boonstra:1998yu}. It also arises in type IIB supergravity as an intersection of D1's and D5's \cite{Cowdall:1998bu}. The solution preserves 16 supersymmetries and has the superisometry group $D(2,1;\alpha)\times D(2,1;\alpha)\times U(1)$. The geometry is supported by NSNS and RR four-form flux of the form
\begin{eqnarray}
H&=&2q\big(\Omega_{AdS_3}+\sqrt\alpha\,\Omega_{S^3_1}+\sqrt{1-\alpha}\,\Omega_{S^3_2}\big)\,,\nonumber\\
F^{(4)}&=&2\hat qe^{-\phi}dx^9\big(\Omega_{AdS_3}+\sqrt\alpha\,\Omega_{S^3_1}+\sqrt{1-\alpha}\,\Omega_{S^3_2}\big)\,,
\end{eqnarray}
where the parameters $q$, $\hat q$ satisfy $q^2+{\hat q}^2=1$. When $q=0$ there is only RR-flux and conversely when $q=1$, $\hat q=0$ there is only NSNS-flux. The free parameter $q$ arises from the freedom to perform an S-duality in the type IIB picture. From the definition of $S$ in (\ref{eq:S}) and $T$ in (\ref{eq:TandUab}) we find
\begin{equation}
S=-4\hat q\mathcal P\Gamma^{0129}\,,\qquad T=-\frac{i}{2}\Gamma^{012}[\hat q\Gamma^9+2q\Gamma_{11}][1-\mathcal P]\,,
\end{equation}
where the Killing spinor projection matrix is
\begin{equation}
\mathcal P=\frac12(1+\sqrt\alpha\,\Gamma^{012345}+\sqrt{1-\alpha}\,\Gamma^{012678})\,.
\end{equation}

Computing the RHS of the Killing spinor equation (\ref{eq:kxi-comm}) one finds
\begin{equation}
M_a\xi=(H_{abc}\Gamma^{bc}\Gamma_{11}+S\Gamma_a)\xi=\hat S\Gamma_a\xi\qquad\text{with}\qquad\hat S=-4\mathcal P\Gamma^{0129'}\,,
%2A_{[0}A_{1]}=32\Gamma_{01}
\end{equation}
where $\Gamma^{9'}$ is defined in (\ref{eq:rotatedgamma}). The dilatino equation in (\ref{eq:SUSYcond}) is obviously satisfied and it is also easy to check the gravitino equation. One finds the radii of curvature to be
\begin{equation}
R_{AdS_3}=1\,,\qquad R_{S^3_1}=\frac{1}{\sqrt\alpha}\,,\qquad R_{S^3_2}=\frac{1}{\sqrt{1-\alpha}}\,.
\end{equation}
Note that for $\alpha=0\,,1$ one $S^3$ decompactifies and the geometry becomes instead $AdS_3\times S^3\times T^4$.

It is easy to verify that conditions (i), (ii) and (iii) hold. Conditions (iv)--(xi) of table \ref{table:2} are also easily verified.

\subsection*{C. $AdS_3\times S^2\times S^3\times T^2$ (and $AdS_3\times S^2\times T^5$)}
This type IIA solutions can be obtained by starting from the same eleven-dimensional $AdS_3\times S^3\times S^3\times T^2$ solution as in the previous case but, instead of reducing on a $T^2$ direction, performing a Hopf reduction on the $S^3$, i.e. viewing $S^3$ as an $S^1$ fibration over $\mathbbm{CP}^1\sim S^2$ and reducing on the $S^1$. The Hopf reduction breaks the supersymmetry down from sixteen to eight supercharges and the superisometry algebra is $D(2,1;\alpha)\times SL(2,\mathbbm R)\times SU(2)\times U(1)^2$. The fluxes take the form
\begin{equation}
H=dx^9\Omega_{S^2}\,,\qquad F^{(2)}=-e^{-\phi}\Omega_{S^2}\,,\qquad F^{(4)}=e^{-\phi}dx^9(\sqrt{1+\alpha}\,\Omega_{AdS_3}-\sqrt\alpha\,\Omega_{S^3})\,.
\end{equation}
Note the presence of both NSNS and RR flux. From the definition of $S$ in (\ref{eq:S}) and $T$ in (\ref{eq:TandUab}) we find
\begin{equation}
S=-2\mathcal P_2\Gamma^{34}\Gamma_{11}\,,\qquad T=\frac{i}{4}\Gamma^{34}\Gamma_{11}[2(1-\mathcal P_1)+(1-\mathcal P_2)]
\end{equation}
where the Killing spinor projection matrix is now a product of two projection matrices
\begin{equation}
\mathcal P=\mathcal P_1\mathcal P_2\,,\qquad\mbox{with}\qquad\mathcal P_1=\frac12(1+\Gamma^9)\,,\qquad
\mathcal P_2=\frac12(1-\sqrt{1+\alpha}\,\Gamma^{340129}\Gamma_{11}-\sqrt\alpha\,\Gamma^{345679}\Gamma_{11})\,.
\end{equation}
The two projection matrices commute and therefore reduce the amount of supersymmetry by a factor of 2 each from 32 to 8. The dilatino equation in (\ref{eq:SUSYcond}) is obviously satisfied and it is also easy to check the gravitino equation. One finds the radii of curvature to be
\begin{equation}
R_{AdS_3}=\frac{2}{\sqrt{1+\alpha}}\,,\qquad R_{S^2}=1\,,\qquad R_{S^3}=\frac{2}{\sqrt\alpha}
\end{equation}
Taking $\alpha=0$ decompactifies the $S^3$ and we obtain the solution $AdS_3\times S^2\times T^5$. Taking instead the limit $\alpha\rightarrow\infty$ gives a highly curved $AdS_3\times S^3$ subspace with the curvature of the $S^2$ remaining finite. Computing the RHS of the Killing spinor equation (\ref{eq:kxi-comm}) one finds
\begin{equation}
M_a\xi=(H_{abc}\Gamma^{bc}\Gamma_{11}+S\Gamma_a)\xi=c_{[a]}\hat S\Gamma_a\xi\qquad\text{with}\qquad\hat S=-2\mathcal P\Gamma^{34}\Gamma_{11}\,,
\end{equation}
with $c_{[a]}$ defined as in (\ref{eq:ca}).

Next we look at the conditions needed for integrability. Conditions (i)--(iii) are easily verified. Since the NSNS-flux cannot be tuned we have $q=1$ which means that conditions (vi$'$) and (ix$'$) in table \ref{table:3} are empty. Conditions (iv$'$) and (viii$'$) are clearly true and the remaining conditions are easily seen to hold after a little bit of algebra.

\subsection*{D. $AdS_3\times S^2\times S^2\times T^3$ (and $AdS_3\times S^2\times T^5$)}
This solution arises in type IIB supergravity as the near-horizon limit of a self-dual configuration of two D5's two NS5's and one D3 \cite{Boonstra:1998yu}. The fluxes of the T-dual type IIA solution are
\begin{eqnarray}
H&=&-\sqrt\alpha\,dx^7\Omega_{S^2_1}+\sqrt{1-\alpha}\,dx^8\Omega_{S^2_2}\,,\nonumber\\
F^{(4)}&=&e^{-\phi}\big(\sqrt\alpha\,dx^9dx^8\Omega_{S^2_1}+\sqrt{1-\alpha}\,dx^9dx^7\Omega_{S^2_2}-\Omega_{S^2_1}\Omega_{S^2_2}\big)\,.
\end{eqnarray}
It is not possible to switch off the NSNS or RR flux (performing S-duality in the IIB picture only leads to a rotation in the $78$-plane). From the definition of $S$ in (\ref{eq:S}) and $T$ in (\ref{eq:TandUab}) we find
\begin{equation}
S=-2\mathcal P_2\Gamma^{3456}\,,\qquad T=\frac{i}{4}\Gamma^{012}[2\mathcal P_2(1-\mathcal P_1)-(1-\mathcal P_2)]\,,
\end{equation}
where the Killing spinor projection matrix is again a product of two commuting projection matrices
\begin{equation}
\mathcal P=\mathcal P_1\mathcal P_2\,,\qquad\mbox{with}\qquad\mathcal P_1=\frac12(1+\Gamma^{0123456})\,,\qquad
\mathcal P_2=\frac12(1+\sqrt\alpha\,\Gamma^{5689}+\sqrt{1-\alpha}\,\Gamma^{3479})\,,
\end{equation}
leaving eight $\xi$'s. The dilatino equation in (\ref{eq:SUSYcond}) is obviously satisfied and it is also easy to check the gravitino equation. One finds the radii of curvature to be
\begin{equation}
R_{AdS_3}=2\,,\qquad R_{S^2_1}=\frac{1}{\sqrt\alpha}\,,\qquad R_{S^2_2}=\frac{1}{\sqrt{1-\alpha}}\,.
\end{equation}
For $\alpha=0\,,1$ one $S^2$ decompactifies and the geometry becomes instead $AdS_3\times S^2\times T^5$.

Computing the RHS of the Killing spinor equation (\ref{eq:kxi-comm}) one finds
\begin{equation}
M_a\xi=(H_{abc}\Gamma^{bc}\Gamma_{11}+S\Gamma_a)\xi=c_{[a]}\hat S\Gamma_a\xi\qquad\text{with}\qquad\hat S=-2\mathcal P\Gamma^{3456}\,,
\end{equation}
with $c_{[a]}$ defined as in (\ref{eq:ca}). The conditions needed for integrability, (i)--(iii) together with the conditions listed in table \ref{table:3}, are easily verified (note that we have $q=1$ so that e.g. $\Gamma^{9'}=-\Gamma_{11}$).

\if 0

IIB:
\begin{equation}
H=-2\sqrt\alpha\,dx^7\frac{1}{2}e^{\hat b}e^{\hat a}\varepsilon_{\hat a\hat b}+2\sqrt{1-\alpha}\,dx^8\frac{1}{2}e^{b'}e^{a'}\varepsilon_{a'b'}\,,\qquad
F^{(3)}=e^{-\phi}(2\sqrt\alpha\,dx^8\frac{1}{2}e^{\hat b}e^{\hat a}\varepsilon_{\hat a\hat b}+2\sqrt{1-\alpha}\,dx^7\frac{1}{2}e^{b'}e^{a'}\varepsilon_{a'b'})\,,\qquad 
F^{(5)}=2e^{-\phi}(\frac{1}{6}dx^8dx^7e^ce^be^a\varepsilon_{abc}+\mbox{dual})
\end{equation}

the projection matrix is a product of two commuting projection matrices
\begin{equation}
\mathcal P=\mathcal P_1\mathcal P_2\,,\qquad\mbox{with}\qquad\mathcal P_1=\frac12(1+i\sigma^2\gamma^{78})\,,\qquad
\mathcal P_2=\frac12(1+\sigma^1(\sqrt\alpha\,\gamma^{012348}+\sqrt{1-\alpha}\,\gamma^{012567}))
\end{equation}

\begin{equation}
S=4\gamma^{012}[\mathcal P_1-\mathcal P_2]\,,\qquad T=i\gamma^{012}[1-2(1-\mathcal P_2)](1-\mathcal P_1)
\end{equation}

\fi

\subsection*{E. $AdS_2\times S^3\times S^3\times T^2$ (and $AdS_2\times S^3\times T^5$)}
This example is very similar to (C). As in that case one starts from the eleven-dimensional $AdS_3\times S^3\times S^3\times T^2$ solution. Viewing $AdS_3$ as an $S^1$ fibration over $AdS_2$ one reduces on the $S^1$ breaking the supersymmetry from 16 down to 8. The fluxes in the IIA solution take the form
\begin{equation}
H=dx^9\Omega_{AdS_2}\,,\qquad F^{(2)}=-e^{-\phi}\Omega_{AdS_2}\,,\qquad F^{(4)}=-e^{-\phi}dx^9\big(\sqrt\alpha\,\Omega_{S^3_1}+\sqrt{1-\alpha}\,\Omega_{S^3_2}\big)\,.
\end{equation}
Again both NSNS and RR flux is present. From the definition of $S$ in (\ref{eq:S}) and $T$ in (\ref{eq:TandUab}) we find
\begin{equation}
S=2\mathcal P_2\Gamma^{01}\Gamma_{11}\,,\qquad T=-\frac{i}{4}\Gamma^{01}\Gamma_{11}[2(1-\mathcal P_1)+(1-\mathcal P_2)]
\end{equation}
where the Killing spinor projection matrix is again a product of two commuting projection matrices
\begin{equation}
\mathcal P=\mathcal P_1\mathcal P_2\,,\qquad\mbox{with}\qquad\mathcal P_1=\frac12(1+\Gamma^9)\,,\qquad
\mathcal P_2=\frac12(1-\sqrt\alpha\,\Gamma^{012349}\Gamma_{11}-\sqrt{1-\alpha}\,\Gamma^{015679}\Gamma_{11})\,.
\end{equation}
The dilatino equation in (\ref{eq:SUSYcond}) is obviously satisfied and it is also easy to check the gravitino equation. One finds the radii of curvature to be
\begin{equation}
R_{AdS_2}=1\,,\qquad R_{S^3_1}=\frac{2}{\sqrt\alpha}\,,\qquad R_{S^3_2}=\frac{2}{\sqrt{1-\alpha}}\,.
\end{equation}
For $\alpha=0\,,1$ one $S^3$ decompactifies and the geometry becomes $AdS_2\times S^3\times T^5$.

Computing the RHS of the Killing spinor equation (\ref{eq:kxi-comm}) one finds
\begin{equation}
M_a\xi=(H_{abc}\Gamma^{bc}\Gamma_{11}+S\Gamma_a)\xi=c_{[a]}\hat S\Gamma_a\xi\qquad\text{with}\qquad\hat S=2\mathcal P\Gamma^{01}\Gamma_{11}\,,
\end{equation}
with $c_{[a]}$ defined as in (\ref{eq:ca}). The conditions needed for integrability, (i)--(iii) together with the conditions listed in table \ref{table:3}, are easily verified (note that we have $q=1$ so that e.g. $\Gamma^{9'}=-\Gamma_{11}$).

\subsection*{F. $AdS_2\times S^2\times S^3\times T^3$ (and $AdS_2\times S^3\times T^5$)}
This solution is similar to (D). It can be realized in type IIB as the near-horizon geometry of a D1-F1-D5-NS5-D3 intersection \cite{Boonstra:1998yu}. The fluxes of the T-dual type IIA solution are
\begin{eqnarray}
H&=&-\sqrt{1+\alpha}\,dx^7\Omega_{AdS_2}+\sqrt\alpha\,dx^8\Omega_{S^2}\,,\nonumber\\
F^{(4)}&=&e^{-\phi}\big(\sqrt{1+\alpha}\,dx^9dx^8\Omega_{AdS_2}+\sqrt\alpha\,dx^9dx^7\Omega_{S^2}-\Omega_{AdS_2}\Omega_{S^2}\big)\,.
\end{eqnarray}
Both NSNS and RR flux is present. From the definition of $S$ in (\ref{eq:S}) and $T$ in (\ref{eq:TandUab}) we find
\begin{equation}
S=2\mathcal P_2\Gamma^{0123}\,,\qquad T=\frac{i}{4}\Gamma^{456}[2\mathcal P_2(1-\mathcal P_1)-(1-\mathcal P_2)]\,,
\end{equation}
where the Killing spinor projection matrix is again a product of two commuting projection matrices
\begin{equation}
\mathcal P=\mathcal P_1\mathcal P_2\,,\qquad\mbox{with}\qquad\mathcal P_1=\frac12(1+\Gamma^{0123456})\,,\qquad
\mathcal P_2=\frac12(1+\sqrt\alpha\,\Gamma^{0179}+\sqrt{1+\alpha}\,\Gamma^{2389})\,.
\end{equation}
The dilatino equation in (\ref{eq:SUSYcond}) is obviously satisfied and it is also easy to check the gravitino equation. One finds the radii of curvature to be
\begin{equation}
R_{AdS_2}=\frac{1}{\sqrt{1+\alpha}}\,,\qquad R_{S^2}=\frac{1}{\sqrt\alpha}\,,\qquad R_{S^3}=2\,.
\end{equation}
For $\alpha=0$ the $S^2$ decompactifies and the geometry becomes instead $AdS_3\times S^2\times T^5$. The opposite limit, $\alpha\rightarrow\infty$, gives a highly curved $AdS_2\times S^2$ subspace with the $S^3$ curvature remaining finite.

Computing the RHS of the Killing spinor equation (\ref{eq:kxi-comm}) one finds
\begin{equation}
M_a\xi=(H_{abc}\Gamma^{bc}\Gamma_{11}+S\Gamma_a)\xi=c_{[a]}\hat S\Gamma_a\xi\qquad\text{with}\qquad\hat S=2\mathcal P\Gamma^{0123}\,,
\end{equation}
with $c_{[a]}$ defined as in (\ref{eq:ca}). The conditions needed for integrability, (i)--(iii) together with the conditions listed in table \ref{table:3}, are easily verified (note that we have $q=1$ so that e.g. $\Gamma^{9'}=-\Gamma_{11}$).

\subsection*{G. $AdS_2\times S^2\times S^2\times T^4$ (and $AdS_2\times S^2\times T^6$)}
This solution can be obtained by dimensional reduction from the eleven-dimensional supergravity solution $AdS_2\times S^2\times S^2\times T^5$ which arises as the near-horizon geometry of an intersection of two M2's and four M5's \cite{Gauntlett:1997pk}. The superisometry group is $D(2,1;\alpha)\times U(1)^4$ and the fluxes take the form
\begin{eqnarray}
H&=&q\big(-dx^8\Omega_{AdS_2}+\sqrt\alpha\,dx^7\Omega_{S^2_1}+\sqrt{1-\alpha}\,dx^6\Omega_{S^2_2}\big)\,,\nonumber\\
F^{(4)}&=&e^{-\phi}\big([dx^7dx^6-\hat qdx^9dx^8]\Omega_{AdS_2}+\sqrt\alpha[dx^8dx^6+\hat qdx^9dx^7]\Omega_{S^2_1}\nonumber\\
&&{}+\sqrt{1-\alpha}[\hat qdx^9dx^6-dx^8dx^7]\Omega_{S^2_2}\big)\,.
\end{eqnarray}
Here $q^2+\hat q^2=1$ and taking $q=0$ turns off the NSNS-flux ($q=1$ does not however turn off the RR-flux). The parameter $q$ arises from the freedom to perform S-duality in the T-dual type IIB solution or performing the dimensional reduction from eleven dimensions at an angle. From the definition of $S$ in (\ref{eq:S}) and $T$ in (\ref{eq:TandUab}) we find
\begin{equation}
S=-2\mathcal P_2\Gamma^{0167}(1+\hat q\Gamma^{6789})\,,\qquad T=-\frac{i}{4}\Gamma^{0167}(1+\Gamma^{678}[2q\Gamma_{11}+\hat q\Gamma^9])(1-\mathcal P_2)\,,
\end{equation}
where the Killing spinor projection matrix is a product of two commuting projection matrices
\begin{equation}
\mathcal P=\mathcal P_1\mathcal P_2\,,\qquad\mbox{with}\qquad\mathcal P_1=\frac12(1+\Gamma^{6789'})\,,\qquad
\mathcal P_2=\frac12(1+\sqrt\alpha\,\Gamma^{012378}+\sqrt{1-\alpha}\,\Gamma^{014568})\,,
\end{equation}
where $\Gamma^{9'}$ appearing in $\mathcal P_1$ is defined in (\ref{eq:rotatedgamma}). The dilatino equation in (\ref{eq:SUSYcond}) is obviously satisfied and it is also easy to check the gravitino equation. One finds the radii of curvature to be
\begin{equation}
R_{AdS_2}=1\,,\qquad R_{S^2_1}=\frac{1}{\sqrt\alpha}\,,\qquad R_{S^2_2}=\frac{1}{\sqrt{1-\alpha}}\,.
\end{equation}
For $\alpha=0\,,1$ one $S^2$ decompactifies and the geometry becomes instead $AdS_2\times S^2\times T^6$.

Computing the RHS of the Killing spinor equation (\ref{eq:kxi-comm}) one finds
\begin{equation}
M_a\xi=(H_{abc}\Gamma^{bc}\Gamma_{11}+S\Gamma_a)\xi=c_{[a]}\hat S\Gamma_a\xi\qquad\text{with}\qquad\hat S=-2\mathcal P\Gamma^{0167}\,,
\end{equation}
with $c_{[a]}$ defined as in (\ref{eq:ca}). The conditions needed for integrability, (i)--(iii) together with the conditions listed in table \ref{table:3}, are not difficult to verify.

\section{Conclusions}\label{sec:conclusions}
In the first part of the paper we determined the form of the superisometry transformations in a general type II supergravity background up to quartic order in $\Theta$. The string action is known in general to the same order \cite{Wulff:2013kga}. We also determined the form of the superisometry Noether current for the string to the same order. The rest of the paper dealt with the special case of symmetric space backgrounds with constant fluxes and dilaton. We postulated a form for the superisometry algebra and a number a extra constraints on the form of the fluxes in the cases with both NSNS and RR flux. We then showed that given this form of the superisometry algebra and the constraints a Lax connection could be constructed at least up to order $\Theta^2$. We then verified that these constraints hold for the backgrounds listed in table \ref{table:1} of the introduction and which arise through intersecting brane constructions. Note that one can also obtain other, more complicated, examples of integrable backgrounds by applying T-dualities and field redefinitions to these. Sometimes S-duality also preserves the integrability as we have mentioned, although this is in general not guaranteed.

There are many interesting questions and possible future directions of work. A very interesting (and difficult) question is to try to classify integrable backgrounds in string theory, see \cite{Stepanchuk:2012xi,Chervonyi:2013eja} and references therein. This is perhaps manageable if one restricts to $AdS$ backgrounds with some amount of supersymmetry and/or constant fluxes and dilaton. A related question which we touched upon briefly is whether one can prove that integrability of the bosonic string extends to the full superstring under certain conditions. We argued that this seems not to happen in general. It would be very nice to find a more general framework to address the question of integrability, something like a generalization of the supercoset construction. Most likely this would have to rely in a crucial way on the kappa symmetry of the string.

It would also be interesting to study other simple backgrounds, for example ones with less supersymmetry. Certainly the easiest class from our point of view are those with constant fluxes. Perhaps these could even be classified. Another approach is to exploit dualities in order to generate new integrable examples.

Finally it would be interesting to study the quantum properties of the integrable strings found here. For example it was recently understood how to incorporate the massless modes that arise in the BMN-limit into the finite gap equations \cite{Lloyd:2013wza}. It would also be very interesting to try to construct the exact S-matrix for these strings, see \cite{Sundin:2013ypa,Bianchi:2013nra,Hoare:2013ida,Engelund:2013fja,Borsato:2013hoa,Abbott:2013kka} for some recent work. It would also be very interesting if one could say something about the CFTs dual to the string backgrounds considered here. Besides their symmetries \cite{Boonstra:1998yu} very little is known about them.

%%%%%%%%%%%%%%%%%%%%%%%%%%%%%%%%%%%%%%%%%%%%%%%%%%%%%%%%%%%%%%%%%%%%%%%%%%

\subsection*{Acknowledgements}
I wish to thank Olof Ohlsson Sax and Arkady Tseytlin for interesting discussions and useful comments. This work was supported by the ERC Advanced grant No.290456 ``Gauge theory -- string theory duality''.

\newpage
\appendix
%{\Large{\bf Appendices}}

\section{Computation of $L^{(1)}L^{(1)}$}\label{sec:L1L1}
Here we will give some details of the calculation of $L^{(1)}L^{(1)}$ the result of which was given in (\ref{eq:L1L1}). Starting from (\ref{eq:L1L1prime}) and using the killing spinor commutator (\ref{eq:xixi-comm}) we find for the terms involving $\nabla_ak_b$
\begin{align}
\frac{i}{8c_{[a]}}
\Big(
&e^ce^d\,\Theta\Gamma_cVW\Gamma^{ab}\hat SW^\dagger V^\dagger\Gamma_d\Theta
+e^ce^d\,\Theta\Gamma_c\Gamma_{11}VW\Gamma^{ab}\hat SW^\dagger V^\dagger\Gamma_{11}\Gamma_d\Theta
\nonumber\\
&-*e^ce^d\,\Theta\Gamma_c\Gamma_{11}VW\Gamma^{ab}\hat SW^\dagger V^\dagger\Gamma_d\Theta
-*e^ce^d\,\Theta\Gamma_cVW\Gamma^{ab}\hat SW^\dagger V^\dagger\Gamma_{11}\Gamma_d\Theta
\Big)\nabla_ak_b\,.
\end{align}
Let us first compute these terms for backgrounds in group I using the constraints listed in table \ref{table:2}. Using (vii), the form of $W_{\text{I}}$ in (\ref{eq:WI}), (viii) and the fact that $c_{[\tilde a]}=1$ we get
\begin{equation}
\frac{i}{64}(\beta^2+q^2\alpha^2)
\left(
e^ce^d\,\Theta\Gamma_cV\Gamma^{ab}[\hat S+\Gamma_{11}\hat S\Gamma_{11}]V^\dagger\Gamma_d\Theta
-*e^ce^d\,\Theta\Gamma_cV\Gamma^{ab}\Gamma_{11}[\hat S+\Gamma_{11}\hat S\Gamma_{11}]V^\dagger\Gamma_d\Theta
\right)\nabla_ak_b\,.
\end{equation}
Using (ix) and the form of $V$ in (\ref{eq:V}) this reduces to
\begin{eqnarray}
\frac{i}{32}(\beta^2+q^2\alpha^2)
\left(
e^ce^d\,\Theta\Gamma_c\Gamma^{ab}S\Gamma_d\Theta
-*e^ce^d\,\Theta\Gamma_c\Gamma^{ab}\Gamma_{11}S\Gamma_d\Theta
\right)\nabla_ak_b\,.
\end{eqnarray}
For backgrounds in group II we get instead, using (ix$'$) of table \ref{table:3}, the form of $W_{\text{II}}$ in (\ref{eq:WII}), (x$'$) and the form of $V$
\begin{eqnarray}
\frac{i}{16c_{[a]}}\hat\beta^2
\left(
e^ce^d\,\Theta\Gamma_c\Gamma^{ab}S\Gamma_d\Theta
-*e^ce^d\,\Theta\Gamma_c\Gamma^{ab}\Gamma_{11}S\Gamma_d\Theta
\right)\nabla_ak_b\,.
\end{eqnarray}

Using (\ref{eq:xixi-comm}) in (\ref{eq:L1L1prime}) we find the following terms involving $k_a$ 
\begin{align}
-\frac{i}{16}
\Big(
&e^ce^d\,\Theta\Gamma_cVW\hat S\Gamma^a\hat SW^\dagger V^\dagger\Gamma_d\Theta
+e^ce^d\,\Theta\Gamma_c\Gamma_{11}VW\hat S\Gamma^a\hat SW^\dagger V^\dagger\Gamma_{11}\Gamma_d\Theta
\nonumber\\
&-*e^ce^d\,\Theta\Gamma_c\Gamma_{11}VW\hat S\Gamma^a\hat SW^\dagger V^\dagger\Gamma_d\Theta
-*e^ce^d\,\Theta\Gamma_cVW\hat S\Gamma^a\hat SW^\dagger V^\dagger\Gamma_{11}\Gamma_d\Theta
\Big)k_a\,.
\end{align}
Using (vii), (ix$'$), the form of $W$, the fact that $\hat S\Gamma^a\hat S=\frac{1}{c_{[a]}}M^a\hat S$ and (iii) this becomes
\begin{align}
-\frac{i}{16c_{[a]}}
\Big(
&e^ce^d\,\Theta\Gamma_cVW^2M^a\hat SV^\dagger\Gamma_d\Theta
+e^ce^d\,\Theta\Gamma_c\Gamma_{11}VW^2M^a\hat SV^\dagger\Gamma_{11}\Gamma_d\Theta
\nonumber\\
&-*e^ce^d\,\Theta\Gamma_c\Gamma_{11}VW^2M^a\hat SV^\dagger\Gamma_d\Theta
-*e^ce^d\,\Theta\Gamma_cVW^2M^a\hat SV^\dagger\Gamma_{11}\Gamma_d\Theta
\Big)k_a\,.
\end{align}
Consider group I first. Using the form of $W_{\text{I}}$ (\ref{eq:WI}) and $V$ (\ref{eq:V}), the form of $M_a$ (\ref{eq:Killing}) and (viii) of table \ref{table:2} we get
\begin{align}
&-\frac{i}{128}(\beta^2+q^2\alpha^2)
\left(
e^ce^d\,\Theta\Gamma_cS\Gamma^a[\hat S-\Gamma_{11}\hat S\Gamma_{11}]\Gamma_d\Theta
+*e^ce^d\,\Theta\Gamma_cS\Gamma^a\Gamma_{11}[\hat S-\Gamma_{11}\hat S\Gamma_{11}]\Gamma_d\Theta
\right)k_a
\nonumber\\
&-\frac{i}{128}(\beta^2+q^2\alpha^2)
\left(
e^ce^d\,\Theta\Gamma_c\Gamma_{ef}\Gamma_{11}[\hat S+\Gamma_{11}\hat S\Gamma_{11}]\Gamma_d\Theta
-*e^ce^d\,\Theta\Gamma_c\Gamma_{ef}[\hat S+\Gamma_{11}\hat S\Gamma_{11}]\Gamma_d\Theta
\right)
H^{efa}k_a
\nonumber\\
&-\frac{iq}{64\hat q}\alpha\beta
\left(
e^ce^d\,\Theta\Gamma_cS\Gamma^{\tilde a9}[\hat S+\Gamma_{11}\hat S\Gamma_{11}]\Gamma_d\Theta
-*e^ce^d\,\Theta\Gamma_cS\Gamma^{\tilde a9}\Gamma_{11}[\hat S+\Gamma_{11}\hat S\Gamma_{11}]\Gamma_d\Theta
\right)
k_{\tilde a}\,.
\end{align}
Using (ix) and (x) the first two terms cancel and, using (xi), we are left with
\begin{eqnarray}
\frac{i}{32}\alpha\beta
\left(
e^ce^d\,\Theta\Gamma_c\Gamma^{\tilde a\tilde b}S\Gamma_d\Theta
-*e^ce^d\,\Theta\Gamma_c\Gamma^{\tilde a\tilde b}S\Gamma_d\Theta
\right)
H_{\tilde a\tilde b\tilde e}k^{\tilde e}\,.
\end{eqnarray}
Next we turn to group II. Using the form of $W_{\text{II}}$ (\ref{eq:WII}) and $V$ (\ref{eq:V}), the form of $M_a$ (\ref{eq:Killing}), $c_{[\hat a]}=2$, (vi$'$) and (vii$'$) of table \ref{table:3} we get
\begin{align}
&\frac{i}{64}\hat\beta
\Big(
e^ce^d\,\Theta\Gamma_c\Gamma_{a'\hat b}(\alpha-\beta\Gamma_{11})[\hat S+\Gamma_{11}\hat S\Gamma_{11}]\Gamma_d\Theta
\nonumber\\
&\qquad\qquad
-*e^ce^d\,\Theta\Gamma_c\Gamma_{a'\hat b}(\alpha-\beta\Gamma_{11})\Gamma_{11}[\hat S+\Gamma_{11}\hat S\Gamma_{11}]\Gamma_d\Theta
\Big)
H^{a'\hat b\hat e}k_{\hat e}
\nonumber\\
&+\frac{i}{128}\hat\beta
\Big(
e^ce^d\,\Theta\Gamma_c(\alpha+\beta\Gamma_{11})S\Gamma^{\hat a}\Gamma_{11}[\hat S-\Gamma_{11}\hat S\Gamma_{11}]\Gamma_d\Theta
\nonumber\\
&\qquad\qquad
+*e^ce^d\,\Theta\Gamma_c(\alpha+\beta\Gamma_{11})S\Gamma^{\hat a}[\hat S-\Gamma_{11}\hat S\Gamma_{11}]\Gamma_d\Theta
\Big)
k_{\hat a}
\nonumber\\
&-\frac{i}{64}\beta\hat\beta
\left(
e^ce^d\,\Theta\Gamma_cS\Gamma^{\tilde a}[\hat S-\Gamma_{11}\hat S\Gamma_{11}]\Gamma_d\Theta
+*e^ce^d\,\Theta\Gamma_cS\Gamma^{\tilde a}\Gamma_{11}[\hat S-\Gamma_{11}\hat S\Gamma_{11}]\Gamma_d\Theta
\right)
k_{\tilde a}
\nonumber\\
&+\frac{i\hat q}{64}\alpha\hat\beta
\Big(
e^ce^d\,\Theta\Gamma_cVS\Gamma^{\tilde a}\Gamma_{11'}\hat SV^\dagger\Gamma_d\Theta
+e^ce^d\,\Theta\Gamma_c\Gamma_{11}VS\Gamma^{\tilde a}\Gamma_{11'}\hat SV^\dagger\Gamma_{11}\Gamma_d\Theta
\nonumber\\
&\qquad\qquad
-*e^ce^d\,\Theta\Gamma_c\Gamma_{11}VS\Gamma^{\tilde a}\Gamma_{11'}\hat SV^\dagger\Gamma_d\Theta
-*e^ce^d\,\Theta\Gamma_cVS\Gamma^{\tilde a}\Gamma_{11'}\hat SV^\dagger\Gamma_{11}\Gamma_d\Theta
\Big)
k_{\tilde a}\,.
\end{align}
The last term vanishes due to (viii$'$) and using (x$'$), (xi$'$) and (xii$'$) this reduces to
\begin{align}
&\frac{i}{32}\hat\beta
\left(
e^ce^d\,\Theta\Gamma_c\Gamma_{a'\hat b}(\alpha-\beta\Gamma_{11})S\Gamma_d\Theta
-*e^ce^d\,\Theta\Gamma_c\Gamma_{a'\hat b}(\alpha-\beta\Gamma_{11})\Gamma_{11}S\Gamma_d\Theta
\right)
H^{a'\hat b\hat e}k_{\hat e}
\nonumber\\
&+\frac{i}{64}\beta\hat\beta
\left(
e^ce^d\,\Theta\Gamma_c\Gamma^{\tilde a\tilde b}S\Gamma_d\Theta
-*e^ce^d\,\Theta\Gamma_c\Gamma^{\tilde a\tilde b}\Gamma_{11}S\Gamma_d\Theta
\right)
R_{[\tilde g]}R_{\tilde a\tilde b}{}^{\tilde e\tilde f}\varepsilon_{\tilde e\tilde f\tilde g}k^{\tilde g}\,.
\end{align}
Summarizing our calculation we have for group I
\begin{align}
L^{(1)}L^{(1)}=&
\frac{i}{32}\alpha\beta
\left(
e^ce^d\,\Theta\Gamma_c\Gamma_{\tilde a\tilde b}S\Gamma_d\Theta
-*e^ce^d\,\Theta\Gamma_c\Gamma_{\tilde a\tilde b}S\Gamma_d\Theta
\right)
H^{\tilde a\tilde b\tilde e}k_{\tilde e}
\nonumber\\
&+\frac{i}{32}(\beta^2+q^2\alpha^2)
\left(
e^ce^d\,\Theta\Gamma_c\Gamma^{ab}S\Gamma_d\Theta
-*e^ce^d\,\Theta\Gamma_c\Gamma^{ab}\Gamma_{11}S\Gamma_d\Theta
\right)\nabla_ak_b\,,
\end{align}
while for group II we have
\begin{align}
L^{(1)}L^{(1)}=&
\frac{i}{32}\hat\beta
\left(
e^ce^d\,\Theta\Gamma_c\Gamma^{a'\hat b}(\alpha-\beta\Gamma_{11})S\Gamma_d\Theta
-*e^ce^d\,\Theta\Gamma_c\Gamma^{a'\hat b}(\alpha-\beta\Gamma_{11})\Gamma_{11}S\Gamma_d\Theta
\right)
H^{a'\hat b\hat e}k_{\hat e}
\nonumber\\
&+\frac{i}{32}\hat\beta^2
\left(
e^ce^d\,\Theta\Gamma_c\Gamma^{\hat a\hat b}S\Gamma_d\Theta
-*e^ce^d\,\Theta\Gamma_c\Gamma^{\hat a\hat b}\Gamma_{11}S\Gamma_d\Theta
\right)\nabla_{\hat a}k_{\hat b}
\nonumber\\
&+\frac{i}{32}\beta^2
\left(
e^ce^d\,\Theta\Gamma_c\Gamma^{\tilde a\tilde b}S\Gamma_d\Theta
-*e^ce^d\,\Theta\Gamma_c\Gamma^{\tilde a\tilde b}\Gamma_{11}S\Gamma_d\Theta
\right)\widetilde{\nabla_{\tilde a}k_{\tilde b}}\,,
\end{align}
where $\widetilde{\nabla_{\tilde a}k_{\tilde b}}$ was defined in (\ref{eq:tildenablak}) and we have used the fact that due to (viii$'$) $\hat\beta k_{\tilde a}=\frac{\beta}{2}k_{\tilde a}$. It is not hard to see that these equations can be written as one, in the form of (\ref{eq:L1L1}).

\end{document}